\documentclass[pdflatex]{sn-jnl}

\usepackage[utf8]{inputenc}
\UseRawInputEncoding

\usepackage{apacite}
\usepackage{colortbl}
\usepackage{float}
\usepackage{supertabular}
\usepackage{rotating}
\usepackage{booktabs}
\usepackage{ragged2e}
\usepackage{longtable}
\usepackage{array}
\usepackage{verbatim} %comments
\usepackage{interval}
\usepackage{amsmath,amssymb,amsfonts}
\usepackage{adjustbox}
\usepackage{textcomp}
\usepackage{multicol}
\usepackage{setspace}
\usepackage{natbib}
\usepackage{soul}
\usepackage{multicol}
\usepackage{multirow}
\usepackage{interval}
\usepackage{adjustbox}
\usepackage{graphicx}%
\usepackage{multirow}%
\usepackage{amsthm}%
\usepackage{mathrsfs}%
\usepackage{manyfoot}%
\usepackage{booktabs}%
\usepackage{listings}%
\usepackage{tabularray}
\usepackage{rotating}
\usepackage{amssymb}
\usepackage{longtable}
\usepackage{array}
\usepackage{colortbl}
\usepackage{amsmath}
\usepackage{graphicx}
\usepackage{ragged2e}
\usepackage{graphicx}

\hypersetup{
  colorlinks   = true, %Colours links instead of ugly boxes
  urlcolor     = blue, %Colour for external hyperlinks
  linkcolor    = blue, %Colour of internal links
  citecolor   = blue %Colour of citations
}
\usepackage{float}

\usepackage{subcaption} % Include the subcaption package for subfigures
\usepackage{setspace}
\usepackage{anyfontsize}

\definecolor{blue}{rgb}{0,0,1}

\begin{document}

% Title must be 250 characters or less.

\title[Predicting Domestic Violence Using Explainable Ensemble Learning]{Predicting Male Domestic Violence Using Explainable Ensemble Learning and Exploratory Data Analysis}

\author*[1,2]{\fnm{Md Abrar} \sur{Jahin}}\email{abrar.jahin.2652@gmail.com}
\author[1]{\fnm{Saleh Akram} \sur{Naife}}\email{naife.iem.kuet@gmail.com}
\author[3]{\fnm{Fatema Tuj Johora} \sur{Lima}}\email{fatema48@student.sust.edu}
\author*[4]{\fnm{M. F.} \sur{Mridha}}\email{firoz.mridha@aiub.edu}
% \author*[5]{\fnm{Jungpil} \sur{Shin}}\email{jpshin@u-aizu.ac.jp}
\author*[5]{\fnm{Md. Jakir} \sur{Hossen}}\email{jakir.hossen@mmu.edu.my}

\affil[1]{\orgdiv{Department of Industrial Engineering and Management}, \orgname{Khulna University of Engineering \& Technology (KUET)}, \orgaddress{\city{Khulna}, \postcode{9203}, \country{Bangladesh}}}
\affil[2]{\orgdiv{Physics and Biology Unit}, \orgname{Okinawa Institute of Science and Technology Graduate University (OIST)}, \orgaddress{\city{Okinawa}, \postcode{904-0412}, \country{Japan}}}
\affil[3]{\orgdiv{Department of Political Studies}, \orgname{Shahjalal University of Science and Technology}, \orgaddress{\city{Sylhet}, \postcode{3114}, \country{Bangladesh}}}
\affil[4]{\orgdiv{Department of Computer Science}, \orgname{American International University-Bangladesh (AIUB)}, \orgaddress{\city{Dhaka}, \postcode{1229}, \country{Bangladesh}}}
\affil[5]{\orgdiv{Department of Robotics and Automation}, \orgname{The Multimedia University}, \postcode{75450 Bukit Beruang}, \orgaddress{\city{Melaka}, \country{Malaysia}}}

% Please keep the abstract below 300 words

\abstract{
Domestic violence is commonly viewed as a gendered issue that primarily affects women, which tends to leave male victims largely overlooked. This study presents a novel, data-driven analysis of male domestic violence (MDV) in Bangladesh, highlighting the factors that influence it and addressing the challenges posed by a significant categorical imbalance of 5:1 and limited data availability. We collected data from nine major cities in Bangladesh and conducted exploratory data analysis (EDA) to understand the underlying dynamics. EDA revealed patterns such as the high prevalence of verbal abuse, the influence of financial dependency, and the role of familial and socio-economic factors in MDV. To predict and analyze MDV, we implemented 10 traditional machine learning (ML) models, three deep learning models, and two ensemble models, including stacking and hybrid approaches. We propose a stacking ensemble model with ANN and CatBoost as base classifiers and Logistic Regression as the meta-model, which demonstrated the best performance, achieving 95\% accuracy, a 99.29\% AUC, and balanced metrics across evaluation criteria. Model-specific feature importance analysis of the base classifiers identified key features influencing their decision-making. Model-agnostic explainable AI techniques, such as SHAP and LIME, provided both local and global insights into the decision-making processes of the proposed model, thereby increasing transparency and interpretability. Statistical validation using paired \( t \)-tests with 10-fold cross-validation and Bonferroni correction (\( \alpha = 0.0036 \)) confirmed the superior performance of our proposed model over alternatives. Our findings challenge the prevailing notion that domestic abuse primarily affects women, emphasizing the need for tailored interventions and support systems for male victims. Our proposed framework improves the analysis, laying the groundwork for developing effective strategies to address this critical issue.}


\maketitle

\section{Introduction}
\label{sec:introduction}
Domestic violence (DV) occurs when one adult in a relationship misuses power to control another. It establishes control and fear in relationships through violence and other forms of abuse. Such violence can take the form of physical assault, psychological abuse, social abuse, financial abuse, or sexual assault~\citep{rakovec-felser_domestic_2014}. DV may involve a range of intentional actions, including sexual, psychological, emotional, and verbal abuse, which negatively affect one's health and perception of self-identity~\citep{alejo_long-term_2014}. When a spouse or intimate partner uses physical violence or aggression to control the behavior of their partner, she or he is committing DV~\citep{thobejane2012patriarchal}. Domestic abuse against men has been described in the literature since the 1950s~\citep{george_riding_1994}. However, there is much less literature on male DV~\citep{tsang_male_2015}. Male victims of domestic abuse have persistently remained hidden within society, often overlooked and marginalized. This oversight stems from a prevailing narrative prevalent in both academic and societal discussions, one that predominantly portrays DV as exclusively perpetrated by men against women, attributing it to entrenched patriarchal structures~\citep{hines_closer_2010}. The existing literature demonstrates the multifaceted nature of abuse experienced by male victims, encompassing physical, emotional, sexual, and verbal forms of violence~\citep{walters_national_2011}. However, this prevailing narrative has contributed to a pervasive lack of awareness surrounding men as victims of domestic abuse, discouraging them from seeking necessary support and assistance~\citep{wright_absent_2016}.

In Bangladesh, societal norms are deeply entrenched in patriarchy and perpetuate female subordination, relegating discussions on DV against men to the sidelines~\citep{mp_gender_2019}. These norms, rooted in gender roles and power dynamics, foster an environment in which male victimization remains largely unacknowledged, cultivating a culture of silence and stigma surrounding men's experiences of abuse~\citep{mp_gender_2019,khan_female_2022}. Newspaper cases often depict severe assaults, murders, and mutilations inflicted by female partners on their male counterparts, illustrating the severity of the issue~\citep{khan_female_2022}. However, many men hesitate to report such abuse because of social apprehensions, and the lack of adequate support structures further complicates their situation~\citep{mp_gender_2019,khan_female_2022}. Societal expectations of male strength and self-reliance inhibit men from seeking help or disclosing their experiences of abuse~\citep{mp_gender_2019,khan_female_2022}. Existing legal frameworks, primarily designed to support female victims, offer limited avenues for men to seek assistance in cases of abuse perpetrated by female partners~\citep{khan_female_2022}.

Our research methodology revolved around a detailed survey and EDA focused on exploring the intricate aspects of DV against men in Bangladesh. EDA serves as a critical tool for uncovering patterns and insights within datasets that illuminate the landscape of MDV. By examining and visualizing data points related to demographic details, relationship dynamics, and behavioral markers, EDA aids in identifying significant correlations and potential indicators of male victimization in domestic settings. For instance, EDA techniques can reveal patterns of abuse across various age groups, educational backgrounds, or income levels, shedding light on the characteristics associated with male victims. It facilitates the identification of outliers or anomalies within the data, which may signify extreme cases or recurring patterns of domestic abuse against men. Through this comprehensive analysis, EDA was instrumental in understanding the nuanced factors contributing to male victimization.

We aim to identify high-risk groups within the population who are particularly vulnerable to MDV. By discerning the most affected demographics, we can design targeted interventions and implement preventive measures to address DV effectively. While traditional statistical methods provide valuable insights, the unprecedented surge in data from internet platforms and electronic health record systems necessitates adopting more advanced analytical approaches. Using ML techniques offers unique advantages in detecting obscure changes and predicting the likelihood of DV from digital text data. This application of predictive analytics is part of a broader scientific movement to address critical global challenges. For instance, advanced forecasting models are now essential tools for ensuring sustainable agriculture and global food security by predicting crop production \cite{mishra_forecasting_2024}. In the same way, in medical diagnostics, hybrid meta-heuristic algorithms are being used to optimize deep learning (DL) models to achieve high accuracy in tasks like oral cancer detection \cite{myriam_advanced_2023}. This approach is also consistent with trends in engineering, where novel bio-inspired optimization algorithms are developed to improve the predictive power of machine learning (ML) models for complex forecasting challenges, such as in wind power engineering \cite{karim2023bioinspired}. Recent work in structural engineering has shown that ML models like PSO-CatBoost, when combined with explainable AI, significantly outperform traditional empirical formulas, validating the power of this framework for complex predictive tasks \cite{khodadadi2024data}. According to \citep{hui_harnessing_2023}, ML is particularly effective in uncovering hidden patterns and predicting outcomes based on complex datasets, making it an indispensable tool in DV research. \citep{rahman_comparative_2023} similarly highlighted the transformative potential of ML by predicting DV vulnerability in Liberian women.

ML algorithms empower us to analyze extensive datasets encompassing demographic information, behavioral patterns, and factors contributing to DV among men. By leveraging these algorithms, we aim to discern critical indicators that distinguish male victims within domestic settings. It allows predictive modeling based on historical data, allowing proactive identification of potential instances of MDV. Still, a major portion of the DV datasets comprises categorical variables, and the datasets are usually imbalanced and short, which poses challenges to the accuracy of the ML models. Categorical data often requires appropriate encoding techniques and may lead to dimensionality issues, potentially affecting the performance of the predictive models. On the other hand, data imbalance and scarcity adversely affect traditional ML algorithms in terms of their ability to learn and converge. 

To improve the interpretability of ML models, our study integrates eXplainable Artificial Intelligence (XAI) methodologies such as Shapley Additive Explanations (SHAP) and Local Interpretable Model-agnostic Explanations (LIME). These methodologies quantify feature importance and provide localized interpretability, reinforcing our understanding of how ML models predict male victimization. By utilizing SHAP and LIME, we aim to gain insights into the decision-making processes of ML models, enabling informed interventions and targeted support for male victims in Bangladesh's societal context. Integrating these methodologies improves the transparency and comprehensibility of predictive models, fostering an effective approach to addressing MDV.

To guide our investigation into the issue of MDV in Bangladesh, we formulated the following research questions:
\begin{enumerate}
\item What are the correlations between personal, societal, and economic factors concerning domestic abuse of men, and how do these factors interact?
\item Which high-risk groups within the population are most affected by MDV, and what are the contextual factors influencing its prevalence?
\item How can ML techniques effectively predict the likelihood of MDV, addressing the issues of excessive categorical features, imbalance, and data scarcity, and what are the most critical factors contributing to its prediction?
\end{enumerate}

Most research on DV has predominantly focused on female victimization, leaving male victimhood underexplored, particularly in Bangladesh. This study challenges traditional gender norms by highlighting cases where women are perpetrators of abuse, underscoring the psychological, physical, and economic abuse faced by men in intimate relationships. While some research has addressed female-perpetrated DV against men in Bangladesh, a significant gap persists in understanding the extent, nature, and socio-cultural factors contributing to MDV. This oversight perpetuates societal stereotypes, leading to a lack of recognition and support for male victims. Advanced methodologies like ML, which hold the potential to identify patterns, predict high-risk groups, and inform interventions, remain underutilized in MDV research. Comprehensive studies that explore the unique socio-cultural dynamics of male victimization in Bangladesh are urgently needed to address these gaps. In this study, we have utilized ML and DL classifier models and ensemble ML techniques to improve predictive accuracy and reliability in identifying MDV patterns. The integration of ensemble models marks a significant contribution, as these methods combine the strengths of multiple algorithms to achieve superior performance. The absence of such advanced approaches in prior MDV research highlights the novelty of our methodology. The lack of XAI techniques in existing studies has limited the transparency and usability of predictive models. By incorporating XAI methods, we address this gap, providing interpretable insights to guide targeted interventions and policy adjustments. 

Our study makes the following contributions-
\begin{enumerate}
\item This study pioneers the application of explainable ML to understand and predict MDV in Bangladesh, expanding the research focus beyond female-centric narratives and introducing a novel computational framework to this critical social issue.
\item Our proposed framework is robust against data with the issues of imbalance, scarcity, and high presence of categorical features, which makes this approach promising for adaptation to other domains. 
\item Through EDA, our study explores the multifaceted nature of MDV across rural, suburban, and urban areas. We examine correlations between age, income, education, and socio-economic factors, identifying high-risk groups and enabling the development of targeted interventions to address DV effectively.
\item Utilizing ensemble techniques and XAI methodologies, our research predicts vulnerabilities associated with MDV while providing transparent and interpretable insights. These findings contribute to actionable policy recommendations, promoting informed decision-making and advocating for tailored strategies to address gender-based violence.
\end{enumerate}

This paper follows a structured approach, comprising a ``\hyperref[sec:Literature Review]{Literature review}" synthesizing relevant state-of-the-art knowledge, a ``\hyperref[sec:Methodology]{Methodology}" detailing the study's design incorporating data collection, preprocessing, EDA, model implementation, and XAI integration, ``\hyperref[sec:Results and Discussion]{Results and discussions}" presenting performance benchmarking and interpretability analysis, and a ``\hyperref[sec:Conclusion]{Conclusions}" summarizing key insights and advocating for further research.

\section{Literature Review}
\label{sec:Literature Review}
While this study introduces a novel computational methodology, it is built upon a substantial foundation of research into MDV. Foundational work has long acknowledged the existence of ``battered husbands"~\citep{steinmetz1997}, and subsequent comprehensive reviews have since established male victimization as a significant and prevalent issue~\citep{drijber2015review}. The field has explored the specific correlates of violence against men~\citep{hines2008correlates}, the unique barriers male victims face when seeking help~\citep{bolen2020help}, and the dynamics of abuse within specific communities~\citep{rowlands2019abuse}. However, despite this critical body of work, the application of advanced, data-driven predictive modeling to this domain remains a nascent area, particularly within the Bangladeshi context. Our study aims to bridge this methodological gap, complementing the existing research with an explainable ML approach.

DV is often framed as a predominantly female issue, yet men are also subject to DV, with prevalence rates ranging from 3.4\% to 20.3\% for domestic physical violence against men~\citep{kolbe_domestic_2020}. In the United States, approximately 300,000-400,000 men are treated violently by their wives or girlfriends~\citep{walker_male_2020}. Results from the 2016/2017 NISVS survey indicated that half of women and over 40\% of men in the United States reported having encountered contact with sexual violence, physical violence, or stalking victimization by an intimate partner at some stage in their lives~\citep{leemis2022national}. In Germany, approximately 17\% of all recorded crimes were related to Intimate partner violence (IPV), with 141.792 victims in 2019~\citep{wormann_males_2021}. Between 2017 and 2019, the total IPV  increased by 2.08\%, while male victims increased by 7.8\%, representing 19\% of IPV cases in 2019~\citep{wormann_males_2021}. Notably, 21\% of men who suffer from partner abuse never disclose their experiences to anyone, which highlights the stigma and challenges they face in seeking help. \citep{walker_male_2020} identified two broad categories of DV: primary and secondary abuse. He included physical violence, sexual violence, controlling behavior, manipulation, domination, and verbal abuse in the primary abuse category. Secondary abuse included the use of children for personal gain and social and legal manipulation. Research on the prevalence and nature of such violence in the neighboring South Asian countries is minimal, except for a few examples~\citep{deshpande2019sociocultural,kumar2012domestic,rathnaweera2017men}. Computer science technology has been utilized to understand DV on several occasions. Amaoui et al. stated that resting-state functional connectivity had been implicated in understanding the neural mechanisms underlying aggression~\citep{amaoui_resting-state_2022}. In Seinfeld et al.'s study, offenders were exposed to a virtual reality scenario depicting DV, shedding light on the dynamics of abusive behavior within intimate relationships~\citep{seinfeld_offenders_2018}. Genuine academic works regarding domestic abuse against men were initiated in the early to mid-1970s~\citep{hines_closer_2010}.

At the current time, the physical violence and mental violence against women have reduced on a larger scale ~\citep{sony_factors_2023}. Little academic work has been conducted on female-perpetrated DV against men in Bangladesh. One study partially captured male victimization in the rural northwestern region of Bangladesh~\citep{khan_female_2022}. Most surveys and case studies investigating DV against women in Bangladesh have been conducted in rural areas and have revealed that rural Bangladeshi women experience an array of physical, sexual, psychological, and emotional abuse inflicted by their male partners~\citep{koenig_womens_2003,wahed_battered_2007}. Obarisiagbon and Omage used both qualitative and quantitative methods to examine the emerging trend in the culture of DV among men in Southern Nigeria~\citep{obarisiagbon2019emerging}. They developed a structured questionnaire to determine whether DV against men exists in Southern Nigeria and the causes of DV against men in Southern Nigeria. Using Bangladesh as a case study, Khan and Arendse introduced key factors contributing to silences, subjugations, and controversies associated with DV against men to present the current knowledge of such violence~\citep{khan_female_2022}. They collected data from secondary sources. Big-Alabo and Itelimo investigated the silent problem of DV in Nigerian men~\citep{big2022ethics}. They examined various sources of DV against men and the importance of ethics in human relationships. The authors used a qualitative analysis method. Karimi employed qualitative methodology to investigate the drivers of DV against men, forms of violence, and coping mechanisms~\citep{karimi_characterizing_2018}. He collected qualitative data through semi-structured interviews and interviews with key informants. He then analyzed the data through content and thematic analysis. Deshpande explored the extent of DV against men and highlighted the causes and effects of unresolved and underreported violence~\citep{deshpande2019sociocultural}. They also discussed legal and sociocultural issues and preventive strategies that can be implemented to reduce violence against men. To collect data, Drijber, Reijnders, \& Ceelen developed a questionnaire consisting of 15 questions~\citep{drijber_male_2013}. Descriptive statistics were used to summarize the data.  Tsui, Cheung, and Leung conducted a qualitative study to identify the issues faced by male victims of partner abuse~\citep{tsui_help-seeking_2010}. Walker et al. used qualitative and quantitative methods to explore men's experiences with female-perpetrated IPV in Australia~\citep{walker_male_2020}. A total of 258 men were selected using the snowball approach. The online survey consisted of open-ended questions. The author aimed to explore the factors that interplay in the decisions of abused men to report, sociocultural barriers to reporting abuse, and post-report social reactions toward victimized men. He conducted a qualitative study and an interpretative phenomenological analysis~\citep{aborisade_report_2024}. Cheung, Leung, \& Tsui explored the phenomenon of men's help-seeking behavior, in general, and among Asian men through descriptive studies~\citep{cheung_asian_2009}. They conducted an internet search. Carmo, Grams, \& Magalhes conducted a study that characterized men as victims of IPV~\citep{carmo_men_2011}. They followed a retrospective analysis of 535 suspected cases of male IPV victims.

Rahman et al. employed ML algorithms such as LightGBM and Random Forest (RF) to predict DV vulnerability in Liberian women, achieving accuracy rates of 81\% and 82\%, respectively~\citep{rahman_comparative_2023}. Hui et al. reviewed 22 studies applying ML techniques to DV detection, highlighting the potential of supervised and unsupervised methods for classification and prediction tasks using diverse data sources like social media and national databases~\citep{hui_harnessing_2023}. Salehi et al. analyzed Persian textual data from social media to classify and predict DV content in Iran, with the Naïve Bayes model achieving the highest accuracy of 86.77\%~\citep{salehi_domestic_2023}. Hossain et al. used ML models, including RF and Logistic Regression (LR), to predict DV in Bangladesh during the COVID-19 pandemic, identifying income and education levels as key factors influencing increased violence~\citep{hossain_prediction_2021}. These studies collectively express the effectiveness of ML in identifying patterns, predicting risks, and analyzing diverse data sources for addressing the multifaceted issue of DV, making it a robust tool for informed intervention strategies. Complementing these direct applications, significant research also focuses on optimizing the ML pipeline itself, with novel bio-inspired algorithms, such as the Waterwheel Plant Optimization Algorithm, being specifically developed for the critical task of feature selection~\citep{alhussan2023waterwheel}. This demonstrates the ongoing effort to improve the accuracy and efficiency of predictive models in complex socio-technical domains. Our work is situated within the growing field of computational social science, where novel techniques are increasingly used to study DV. As comprehensively reviewed by Kouzani~\citep{kouzani2023technological}, this includes a wide range of technological approaches, from analyzing data on digital platforms to using ambient sensors. Specific studies, for instance, have successfully applied ML to detect DV-related discussions on social media platforms like Twitter~\citep{saleem2023detecting}.

Literature on DV has predominantly focused on female victims, neglecting male victims, particularly in the context of Bangladesh. Although studies have explored DV against women in rural Bangladesh, research on female-perpetrated DV against men remains scarce. Existing literature from countries such as Nigeria and Australia has examined the emerging trend of DV against men by employing qualitative and quantitative methodologies to uncover its prevalence, causes, and effects. Still, there is a notable absence of studies integrating ML techniques to analyze MDV data, especially in underrepresented regions. Although qualitative studies have provided valuable insights into the experiences of male victims, the application of ML in this domain remains largely underexplored. So, a significant gap exists in the literature regarding integrating ML methods for investigating MDV, particularly in regions where research on this topic is limited, such as Bangladesh. Also, most of these datasets are predominantly categorical, characterized by significant class imbalance and inadequate sample sizes, which pose challenges for training ML models to effectively learn hidden patterns. To address these limitations, it is essential to develop robust models capable of handling imbalanced data, capturing complex relationships, and learning effectively from limited samples.

\section{Methodology}\label{sec:Methodology}
\begin{figure*}[!ht]
\centering
\includegraphics[width=\linewidth]{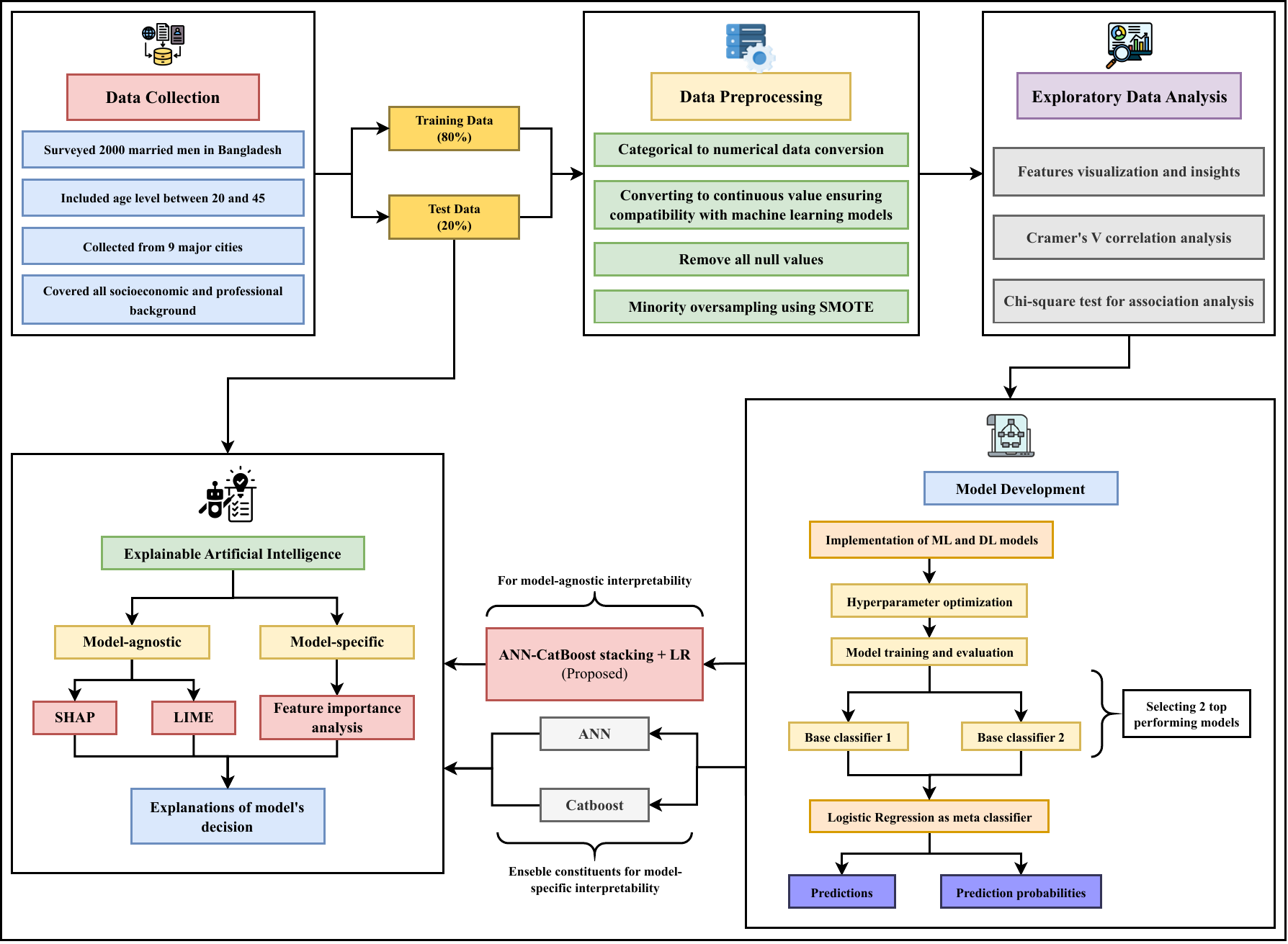}
\caption{Methodological framework illustrating the techniques utilized in data collection, data preprocessing, ML-DL model development, and model interpretability.}
\label{fig:framework}
\end{figure*}
Our study focused on the comprehensive exploration of MDV in Bangladesh. As illustrated in Fig. \ref{fig:framework}, we began by collecting and preprocessing data, which set the stage for a thorough EDA. This analytical phase revealed important patterns and correlations within the dataset, providing insights into the societal context surrounding experiences of male abuse. Using ML methodologies, we developed predictive models to investigate the dynamics of abuse further. We evaluated various models, from traditional approaches to DL techniques, and assessed their performance to select the three most effective models. Through ensemble stacking, we combined two or three top-performing base classifiers with LR as a meta-classifier, ultimately proposing a robust model that effectively manages imbalanced, categorical, and short datasets. We integrated both model-specific and model-agnostic XAI techniques to shed light on the decision-making processes of our proposed model and its base classifiers, ensuring transparency and interpretability throughout the analysis.

\subsection{Data Collection}
The dataset was collected through a survey conducted from January 28, 2023, to September 25, 2023. The main focus of the questions in the survey is outlined in Table~\ref{tab:variables}. The survey instrument was designed by adapting questions from established and validated instruments in the DV literature to confirm content validity. It was administered online to maximize participant anonymity and mitigate response biases, such as social desirability, which are common concerns in research on sensitive topics. Responses were obtained from 2000 participants residing in nine major cities of Bangladesh, including Dhaka, Chittagong, and Sylhet, focusing on professionals across various fields. The survey targeted individuals currently married or in domestic partnerships and included 24 questions exploring demographics, monthly income, abuse experiences, and attitudes toward legal measures. Most questions required binary (yes/no) responses, ensuring quantitative data collection. Confidentiality and anonymity were strictly maintained, adhering to the ethical guidelines of the Declaration of Helsinki. Participants provided informed consent after being briefed on the study's purpose and their rights. The research was approved by the Office of the Director, Research \& Extension Center, Khulna University of Engineering \& Technology, under authorization code KUET/DRE/2023/27(1).

\subsection{Dataset Description}

\begin{table*}[!ht]
\centering
\caption{Description of the features used in our dataset.}
\label{tab:variables}
\resizebox{\linewidth}{!}{%
\begin{tabular}{p{6cm}p{9cm}c}
\toprule
\textbf{Feature} & \textbf{Description} & \textbf{Variable Type} \\ \midrule
Residence Type & Type of residence such as ``rural," ``urban," or ``suburban" & Categorical \\
Place of Residence & Specific city or area of residence & Categorical \\
Age & Age of the respondent & Continuous \\
Profession & Occupation or profession of the respondent & Categorical \\
Level of Education & Educational attainment level of the respondent & Categorical \\
Family Type & Type of family structure like ``single family" or ``joint family" & Categorical \\
Family Having Children & Indicator whether the family has children or not & Binary \\
Head of the Family & Indicator if the respondent is the head of the family or not & Binary \\
Monthly Income & Income range of the respondent per month & Categorical \\
Monthly Income Percentage Allocated for Wife & Percentage of monthly income allocated for the wife & Categorical \\
First Torture Experience & Timeframe of the first experience of torture & Categorical \\
First Type of Torture & Nature or type of the first torture experience & Categorical \\
Current Form of Abuse & Type of abuse currently experienced & Categorical \\
Place of Most Abuse & Location where most abuse occurs & Categorical \\
Type of Marriage & Indicator if the respondent's marriage is ``love marriage" or ``arranged marriage" & Categorical \\
Spouse Divorced Before the Current Marriage & Indicator if the spouse was previously divorced & Binary \\
Duration of Current Marriage & Duration of the current marriage in years & Categorical \\
Extramarital Affairs? & Indicator of the spouse involved in extramarital affairs & Categorical \\
Level of Satisfaction with Marital Life & Satisfaction level with marital life & Categorical \\
Spouse Physically Capable of Having Sex & Indicator of the spouse who is physically capable of having sex & Categorical \\
Level of Satisfaction with Sexual Relationship Patterns & Satisfaction level with sexual relationship patterns & Categorical \\
Do you want Bangladesh to pass a male torture law? & Opinion on passing a male torture law in Bangladesh & Categorical \\ \bottomrule
\end{tabular}
}
\end{table*}

Table~\ref{tab:variables} outlines the variables included in our dataset on MDV in Bangladesh and categorizes them based on their types. Most of the features are categorical, representing various socio-demographic characteristics such as residence type, profession, education level, family structure, and marital satisfaction. The binary variables include family having children, indicators of spousal divorce, and the head of the family. 

\begin{figure*}[!ht]
    \centering
    \includegraphics[width=0.6\linewidth]{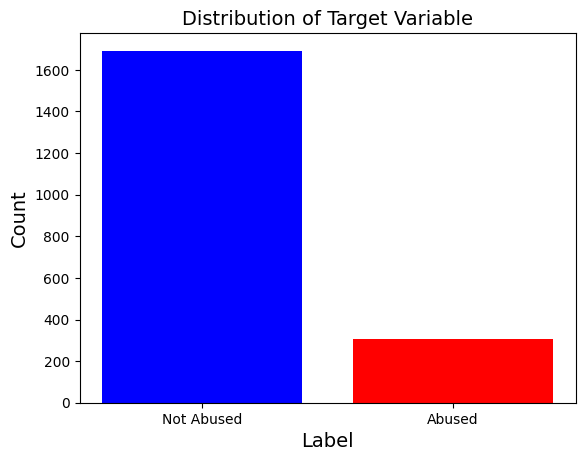}
    \caption{Class distribution of the imbalanced dataset used in our study.}
    \label{fig:17}
\end{figure*}

An important issue encountered in our dataset was the significant class imbalance regarding instances of MDV. Out of the 2000 observations collected, approximately 308 instances (about 15.4\%) were identified as abuse cases, while the remaining 1692 instances (roughly 84.6\%) were classified as non-abuse cases. This class imbalance poses a substantial challenge for predictive modeling, as it can lead models to over-predict the majority class, resulting in a reduced ability to effectively identify cases of abuse. Fig.~\ref{fig:17} illustrates this pronounced imbalance, emphasizing the dominance of non-abuse instances in the dataset.

\subsection{Survey Instrument Validation}
To ensure the quality and interpretability of our data, we validated the survey instrument using standard psychometric techniques, assessing both reliability and construct validity.

\subsubsection{Reliability}
To evaluate the internal consistency of key items, we employed Cronbach's alpha. This test measures the extent to which a set of items is interrelated, thus assessing its capacity to measure a single, unidimensional construct. We constructed a scale for the \textbf{Type of Abuse Experienced}, which included four distinct items: physical, mental, verbal, and sexual abuse. A respondent was coded as having experienced a specific type of abuse if it was mentioned in either the `First type of torture' or `Current form of abuse' fields. The analysis of this 4-item scale yielded a Cronbach's alpha of \textbf{0.887}. A value of this magnitude is considered excellent, far exceeding the commonly accepted threshold of 0.70. This result provides strong evidence that the items measuring the different facets of DV are highly consistent and reliably capture the underlying construct.

\subsubsection{Construct Validity}
To assess the survey's construct validity, we conducted an Exploratory Factor Analysis (EFA). This statistical technique is used to identify the underlying latent structure of a set of variables. The suitability of our data for this analysis was first confirmed. The Kaiser-Meyer-Olkin (KMO) measure of sampling adequacy was 0.505, which is acceptable for factor analysis. Furthermore, Bartlett's Test of Sphericity was highly significant ($p < 0.001$), confirming that the correlations between the variables were sufficiently strong for EFA. The EFA was performed on a subset of eight theoretically related variables. Using a Varimax rotation, a clear three-factor structure emerged, as detailed in Table \ref{tab:factor_loadings}.

\begin{table}[!ht]
\centering
\caption{Factor Loadings from EFA with Varimax Rotation. Significant loadings (absolute value $> 0.50$) are highlighted in \textbf{bold}.}
\label{tab:factor_loadings}
\begin{tabular}{@{}lccc@{}}
\toprule
\textbf{Variable} & \textbf{Factor 1} & \textbf{Factor 2} & \textbf{Factor 3} \\ \midrule
Level of satisfaction with marital life & \textbf{0.992} & 0.016 & -0.101 \\
Current form of abuse & 0.022 & \textbf{0.997} & -0.027 \\
Profession & 0.090 & 0.045 & \textbf{0.778} \\
Monthly income & -0.064 & -0.022 & \textbf{-0.570} \\
Duration of current marriage & 0.057 & -0.005 & 0.048 \\
Extramarital affairs? & -0.010 & 0.034 & 0.052 \\
Level of education & 0.022 & 0.000 & 0.028 \\
Type of marriage & -0.004 & 0.016 & -0.035 \\ \bottomrule
\end{tabular}
\end{table}

The extracted factors are interpretable and align with theoretical expectations:
\begin{enumerate}
    \item \textbf{Factor 1 (Marital Satisfaction):} This factor is almost exclusively defined by the `Level of satisfaction with marital life' variable (loading = 0.992), distinctly representing the construct of relationship quality.
    \item \textbf{Factor 2 (Nature of Current Abuse):} With a near-perfect loading from the `Current form of abuse' variable (0.997), this factor isolates the dimension of the active abuse being experienced.
    \item \textbf{Factor 3 (Socioeconomic Profile):} This factor captures the socioeconomic status of the respondents, with strong loadings from `Profession' (0.778) and `Monthly income' (-0.570).
\end{enumerate}

These results support the internal coherence and construct validity of the instrument, affirming its utility for modeling MDV in our context.

\subsection{Experimental Setup}
In our experimental setup, we utilized Google Colab, a cloud-based platform equipped with an NVIDIA Tesla T4 GPU, to perform high-performance computations. Python 3.10 served as the programming environment, complemented by key libraries. \textit{NumPy (v1.26.4)} and \textit{Pandas} facilitated numerical computations and data manipulation, while \textit{scikit-learn (v1.5.2)} supported data preprocessing, model evaluation, and metric analysis. \textit{TensorFlow} developed and trained DL models, enabling flexible architecture design and optimization. \textit{Matplotlib} and \textit{Seaborn} were also used for data visualization, and \textit{Joblib} ensured efficient model saving and loading.

\subsection{Exploratory Data Analysis}
\subsubsection{Features Visualization and Insights}
%02
\begin{figure*}[!ht]
    \centering
    \includegraphics[width=1\linewidth]{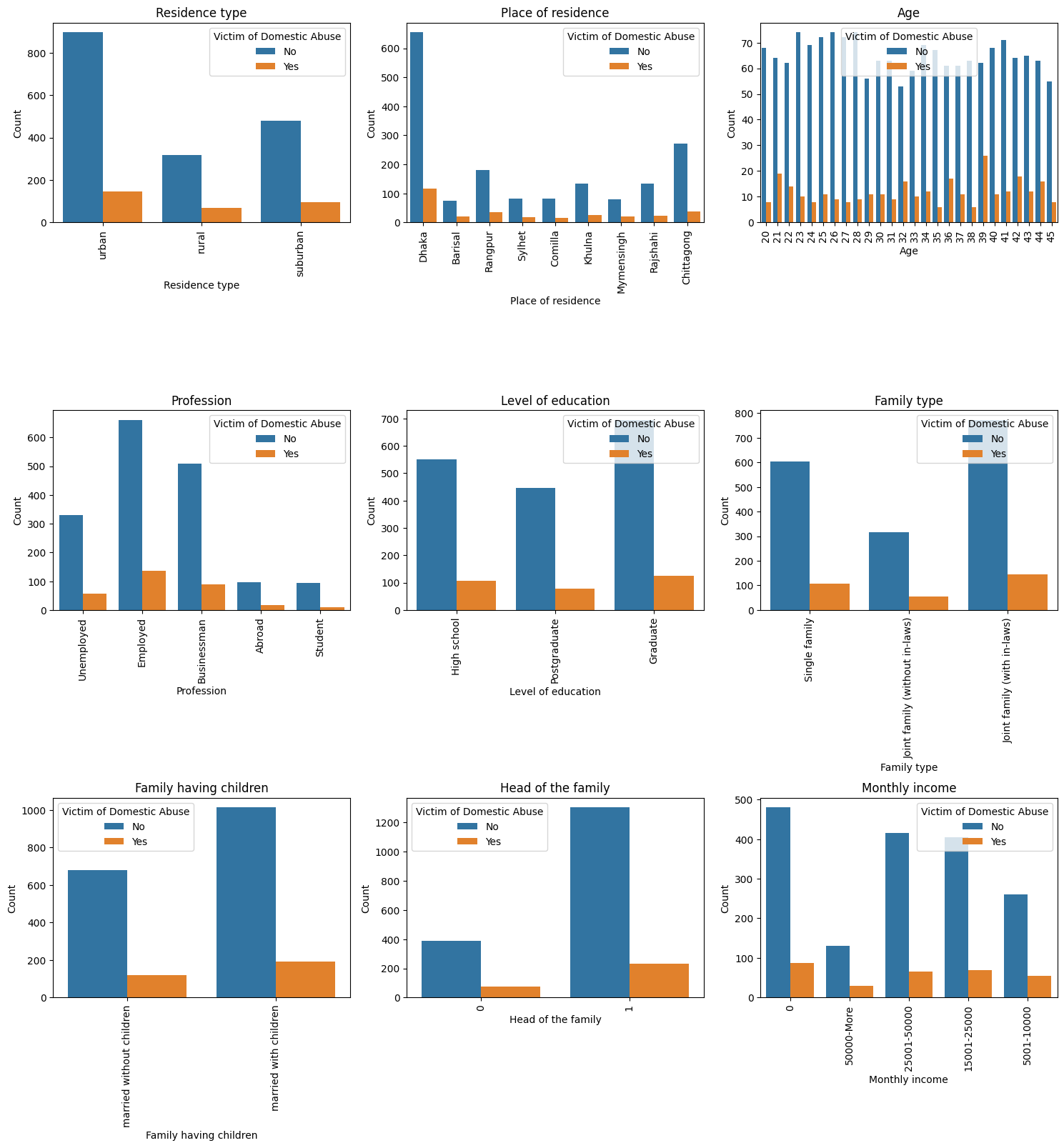}
    \caption{Frequency distribution of demographics and socio-economic factors of the individuals facing MDV.}
    \label{fig:1}
\end{figure*}

Fig.~\ref{fig:1} provides a comprehensive overview of MDV's distribution across demographic and socio-economic categories. Urban areas, particularly Dhaka, report the highest MDV prevalence, likely because of greater population density and reporting accessibility. That said, rural and suburban regions also show significant proportions, indicating that MDV is pervasive across all areas. MDV cases are evenly distributed across age groups, with a slight increase in individuals in their late 20s and early 30s, suggesting that MDV is more linked to life stages and relationship dynamics than age alone. Among professions, unemployed individuals are the most vulnerable, with students and individuals working abroad showing the lowest MDV occurrences. Financial dependency and socio-economic stressors appear to impact vulnerability significantly. High school-educated individuals exhibit the highest prevalence of MDV, while those with graduate and postgraduate education show lower rates, underscoring the protective role of higher education. Joint families with in-laws report the highest MDV incidence, followed by joint families without in-laws, highlighting the complexities introduced by extended family dynamics. Families with children report lower MDV prevalence, possibly because of societal perceptions of parental roles or the mitigating presence of children. Individuals with no income are most vulnerable, with MDV prevalence decreasing as monthly income increases, and a higher percentage of income allocated to wives correlates with reduced MDV.
%03
\begin{figure*}[!ht]
    \centering
    \includegraphics[width=1\linewidth]{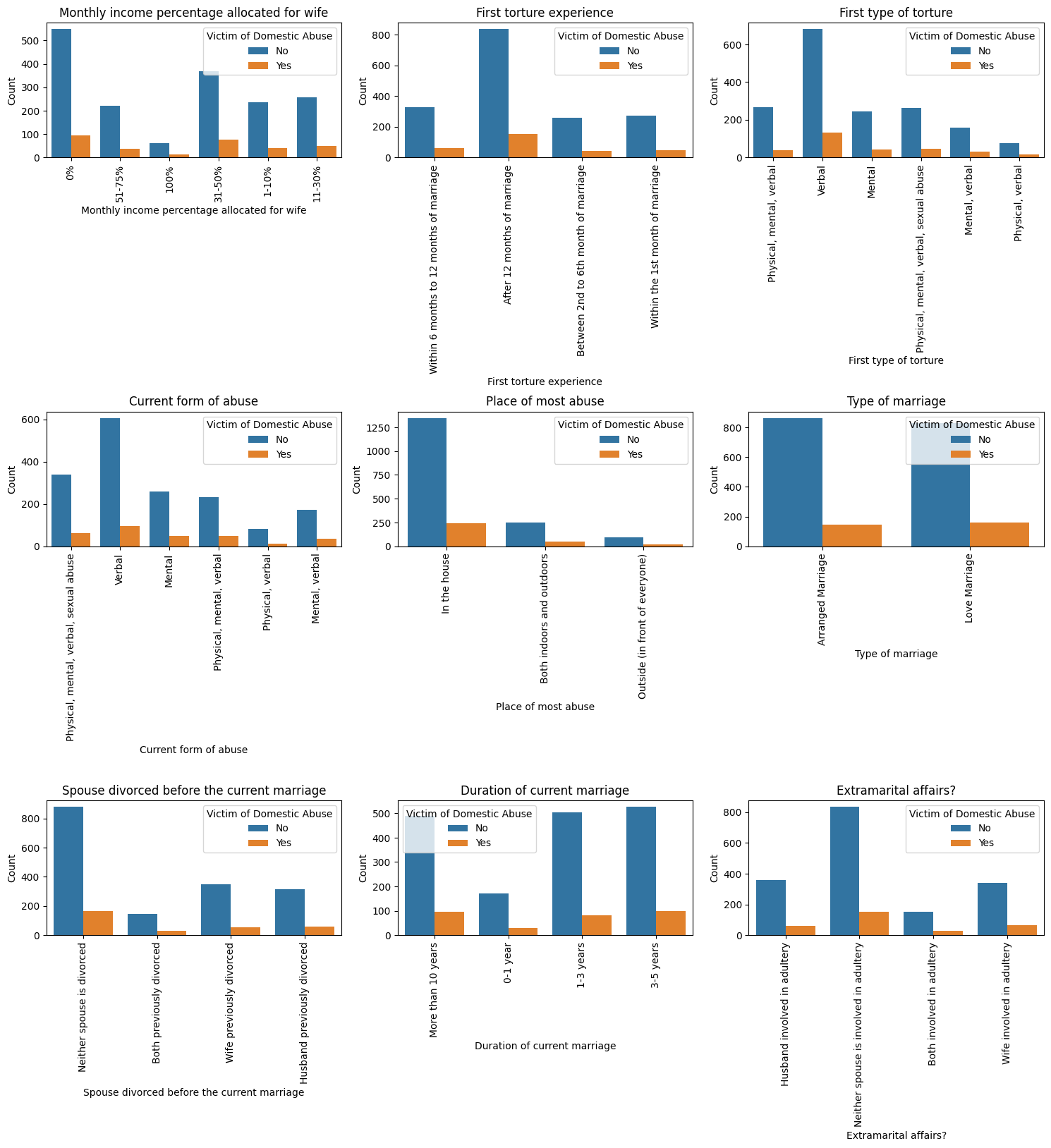}
    \caption{Frequency distribution of economic, familial, and abuse characteristics of the individuals facing MDV.}
    \label{fig:2}
\end{figure*}

Fig.~\ref{fig:2} offers insights into economic and abuse-related factors. MDV is most prevalent when no percentage of the monthly income is allocated to wives, with prevalence decreasing progressively as financial contributions increase. First-time abuse predominantly occurs within the first 12 months of marriage, emphasizing the need for interventions during early marital stages. Verbal abuse is the most common form, often accompanied by mental and physical abuse, highlighting the psychological toll of MDV. Most abuse occurs within households, underscoring the concealed nature of MDV, while a smaller proportion occurs in public settings. Arranged marriages and marriages involving previously divorced spouses show higher MDV rates, reflecting compatibility issues and societal stigma. The highest MDV prevalence occurs within 0–1 year of marriage, gradually declining over time. Extramarital affairs, especially involving husbands, strongly correlate with MDV, emphasizing the destabilizing effect of trust violations.
%04
\begin{figure*}[!ht]
    \centering
    \includegraphics[width=1\linewidth]{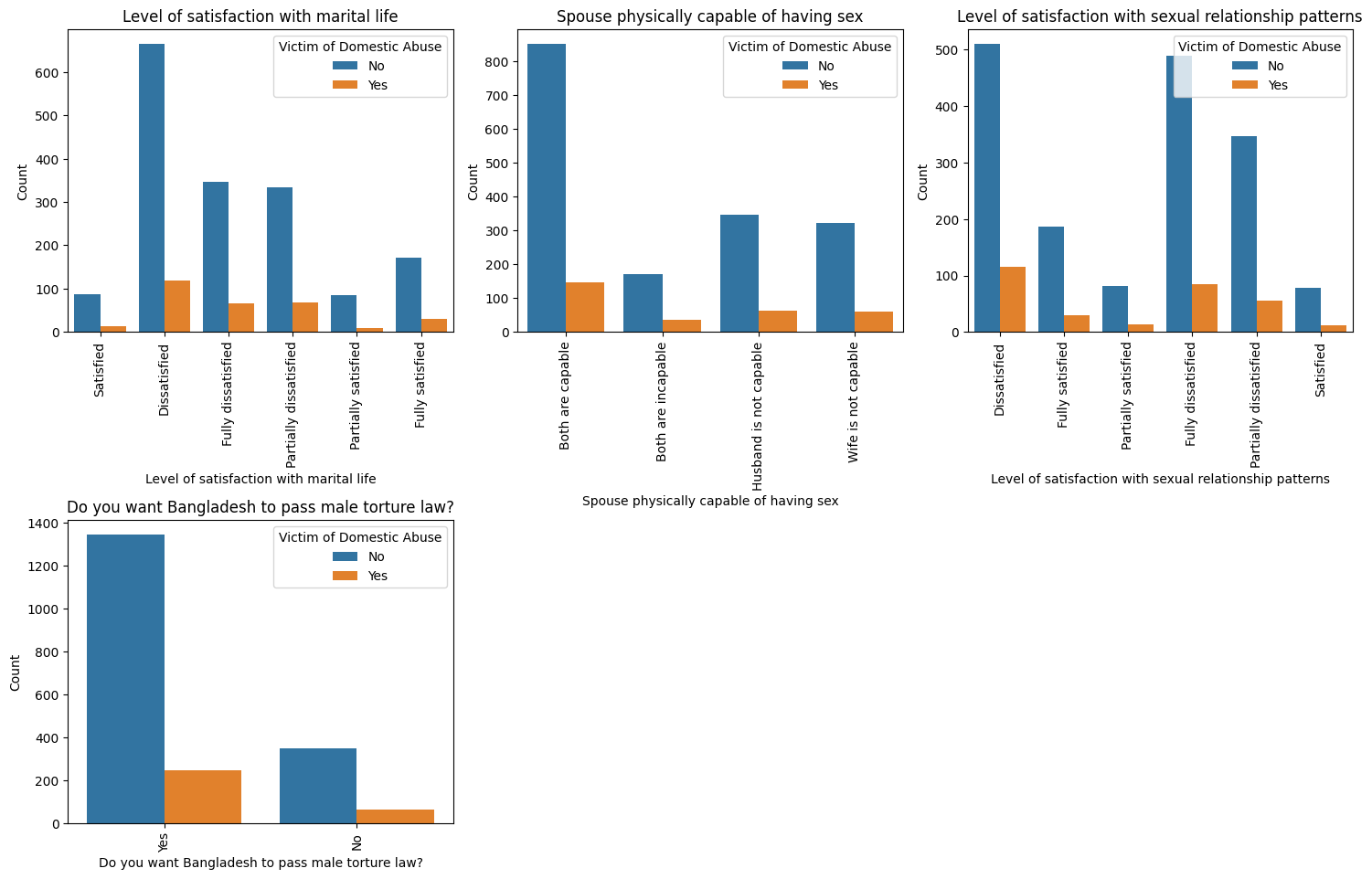}
    \caption{Frequency analysis of features related to abuse, marriage, and other affairs for abused and not-abused individuals.}
    \label{fig:3}
\end{figure*}

Fig.~\ref{fig:3} explores the interplay between marital satisfaction, physical compatibility, and perceptions of legislative reforms. Dissatisfied individuals report significantly higher MDV rates, while full satisfaction with marital life correlates with the lowest MDV prevalence. MDV is most prevalent when both spouses are physically incapable, highlighting the strain of health issues on marital harmony. Higher sexual satisfaction correlates with reduced MDV, emphasizing the importance of open communication and mutual understanding in intimate relationships. Strong public support exists for introducing a male-focused torture law in Bangladesh, reflecting societal demand for gender-specific violence prevention measures.

\begin{figure*}[!ht]
  \centering
    \includegraphics[width=\linewidth]{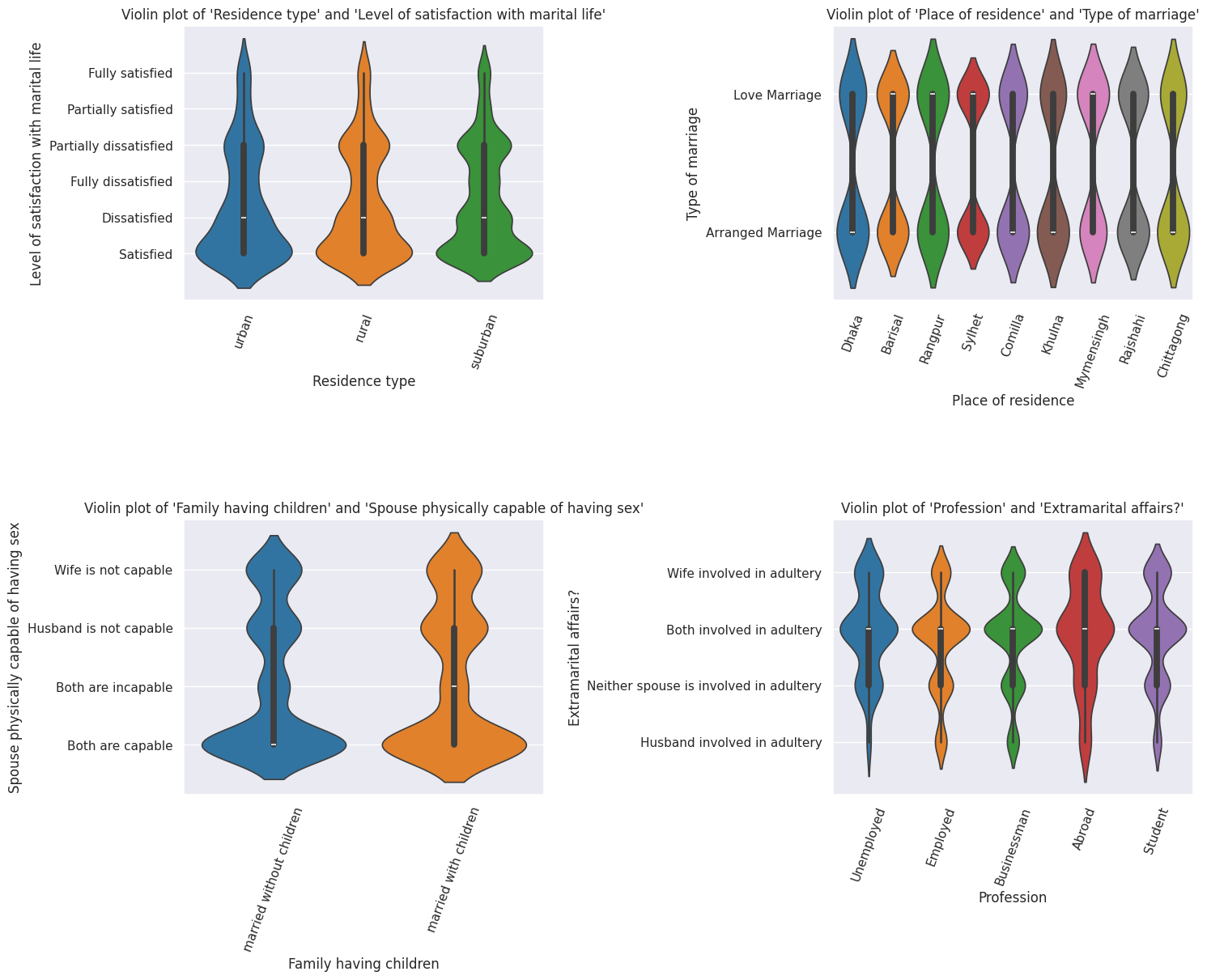}
    \caption{The exploratory violin plots illustrate the relationships between key socio-demographic factors and marital dynamics. The plots include residence type with marital satisfaction (upper left), place of residence versus type of marriage (upper right), age compared to satisfaction with sexual relationships (lower left), and profession with extramarital affairs (lower right).}
    \label{fig:55}
\end{figure*}

Fig.~\ref{fig:55} presents violin plots illustrating the intricate relationships between socio-demographic factors and MDV dynamics in Bangladesh. The relationship between residence type (rural, urban, suburban) and marital satisfaction reveals a wide distribution in urban areas, where dissatisfaction levels are more pronounced, suggesting the influence of urban stressors on marital harmony. In contrast, rural and suburban regions show narrower distributions with fewer dissatisfied individuals, indicating relative stability in these settings.  The correlation between place of residence and type of marriage highlights cultural shifts. Urban areas like Dhaka and Chittagong display a higher prevalence of love marriages, whereas traditional norms dominate in rural regions like Sylhet and Barisal, where arranged marriages are more common.  Family structure and physical capability in sexual relationships show that families without children exhibit greater densities where one or both spouses are physically incapable. In contrast, families with children predominantly report both partners as capable, emphasizing how unmet expectations around family-building may heighten marital tensions.  
Profession and involvement in extramarital affairs reveal that individuals working abroad or in business show higher densities for infidelity, while students and unemployed individuals display lower prevalence. These patterns highlight the role of profession-related factors like prolonged absences or financial independence in straining marital dynamics. 

%07
\begin{figure*}[!ht]
  \centering
    \includegraphics[width=\linewidth]{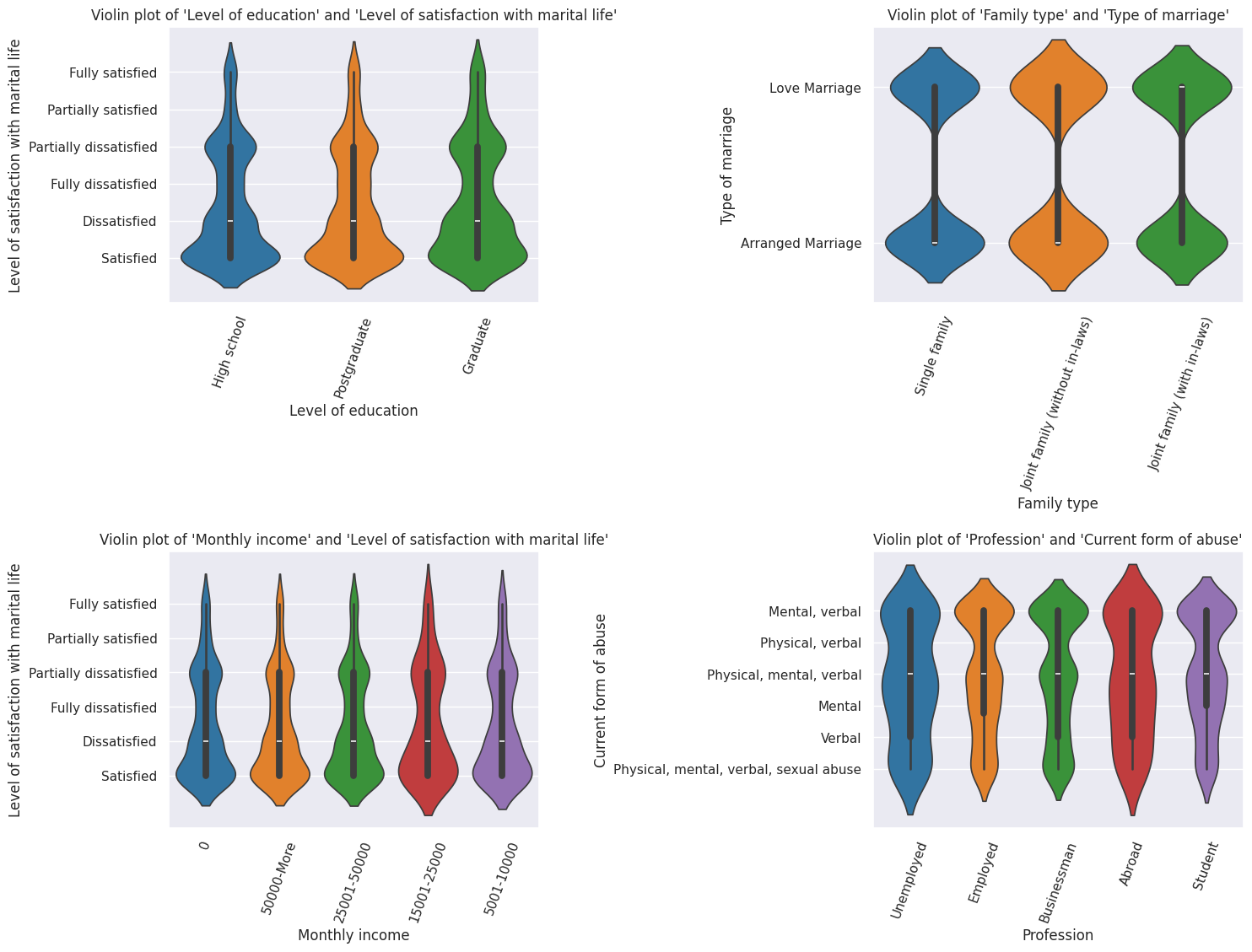}
    \caption{Exploratory violin plots illustrating the complex relationships between key socio-demographic factors and relational dynamics include: (upper left) educational attainment with marital satisfaction, (upper right) family structure compared with marriage type, (lower left) the distribution of monthly income and its effect on marital satisfaction, and (lower right) professional occupation contrasted with patterns of abuse.}
    \label{fig:5}
\end{figure*}

Fig.~\ref{fig:5} provides insights into socio-economic and demographic factors affecting MDV dynamics. Higher education levels are associated with greater marital satisfaction, while individuals with only a high school education exhibit broader spreads, including higher dissatisfaction levels. This suggests that education may mitigate MDV risks by fostering resilience. Family type influences marriage dynamics, with joint families showing a strong association with arranged marriages, reflecting traditional norms that can exacerbate MDV. Single-family structures, by contrast, display a higher prevalence of love marriages, indicating fewer external pressures in marital decision-making. Monthly income correlates with marital satisfaction, peaking in middle-income groups (15,001--250,000 BDT). Satisfaction declines slightly at higher income levels, indicating that financial stability alone does not guarantee marital harmony. The analysis of profession and current forms of abuse reveals that unemployed individuals and those working abroad experience higher rates of verbal and mental abuse, while businessmen exhibit a prevalence of physical and multi-dimensional abuse, reflecting profession-specific stressors influencing MDV. 

\begin{figure*}[!ht]
  \centering
    \includegraphics[width=\linewidth]{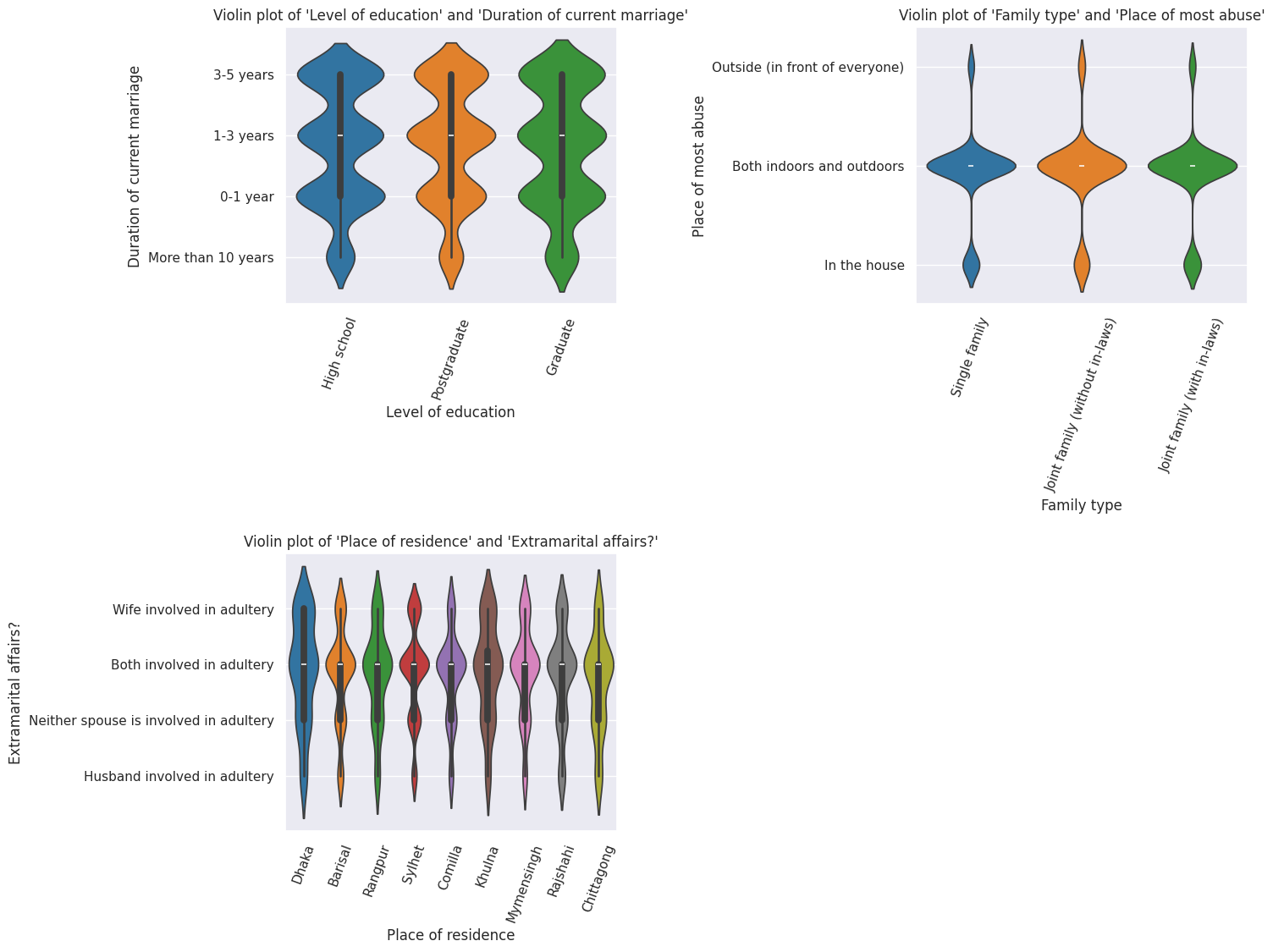}
    \caption{Exploratory violin plots depicting the nuanced interplay between educational attainment versus current marriage duration (upper left), family type versus the place of most abuse (upper right), and the presence of children versus spouses physically capable of sexual intercourse (lower left).}
    \label{fig:6}
\end{figure*}

Fig.~\ref{fig:6} explores the impact of education, family type, and geographical location on MDV. Education level correlates with marriage duration, as individuals with only a high school education display broader distributions, including longer marriages. Higher education is linked to shorter durations, reflecting socio-economic influences on marital stability. Joint families face heightened abuse levels, both indoors and outdoors, whereas single-family structures report fewer public abuse cases, emphasizing the impact of shared living dynamics. Urban areas like Dhaka show higher rates of extramarital affairs, reflecting the complexities of urban lifestyles, while rural areas exhibit minimal involvement, influenced by cultural conservatism. The patterns reiterate how family structure amplifies vulnerabilities to MDV in shared living environments.
%09
\begin{figure*}[!ht]
  \centering
    \includegraphics[width=\linewidth]{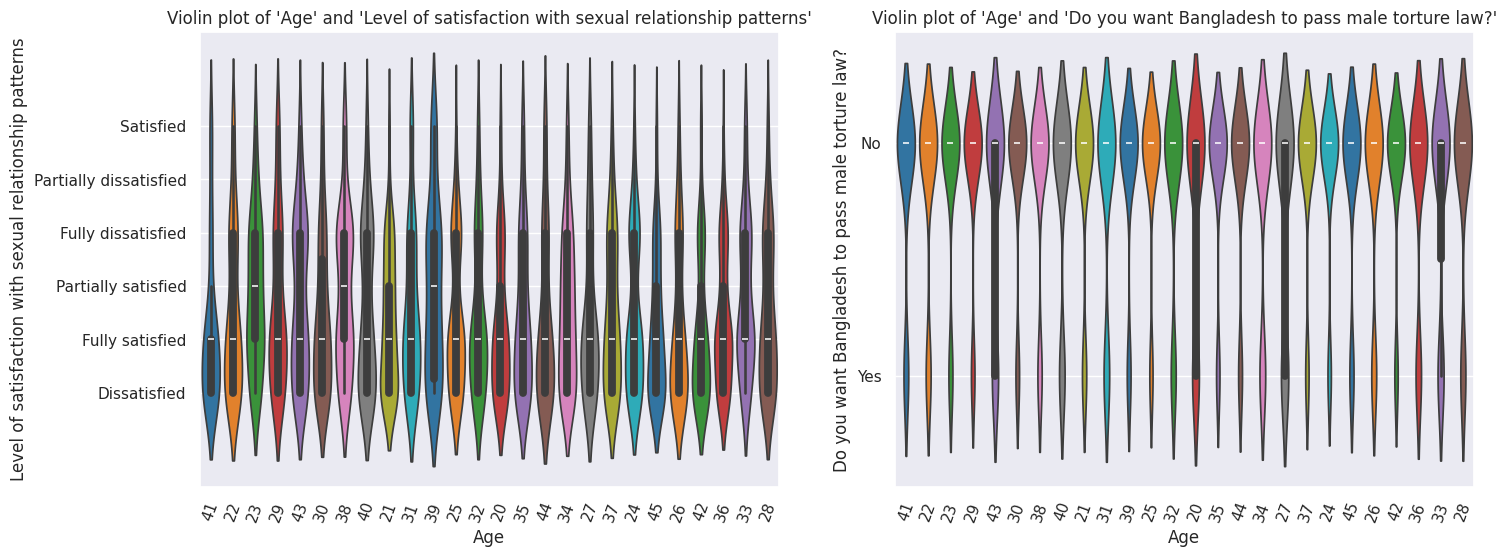}
    \caption{Violin plots depicting the relationship between age and level of satisfaction with sexual relationship patterns, as well as age and legal action against male torture (from left to right).}
    \label{violin4}
\end{figure*}

Fig.~\ref{violin4} focuses on age-related dynamics in MDV. Younger individuals (20--30 years) exhibit wider variability in sexual satisfaction, reflecting relational volatility, while older respondents (40+) display narrower distributions, indicating stability in perceptions.  Age also influences support for male torture law reform. A strong consensus for reform emerges across all age groups, with older individuals displaying consistent support, likely reflecting their accumulated life experiences. These violin plots offer a nuanced understanding of the socio-demographic and relational dynamics shaping MDV in Bangladesh.

\begin{figure*}[!ht]
    \centering
    \includegraphics[width=\textwidth]{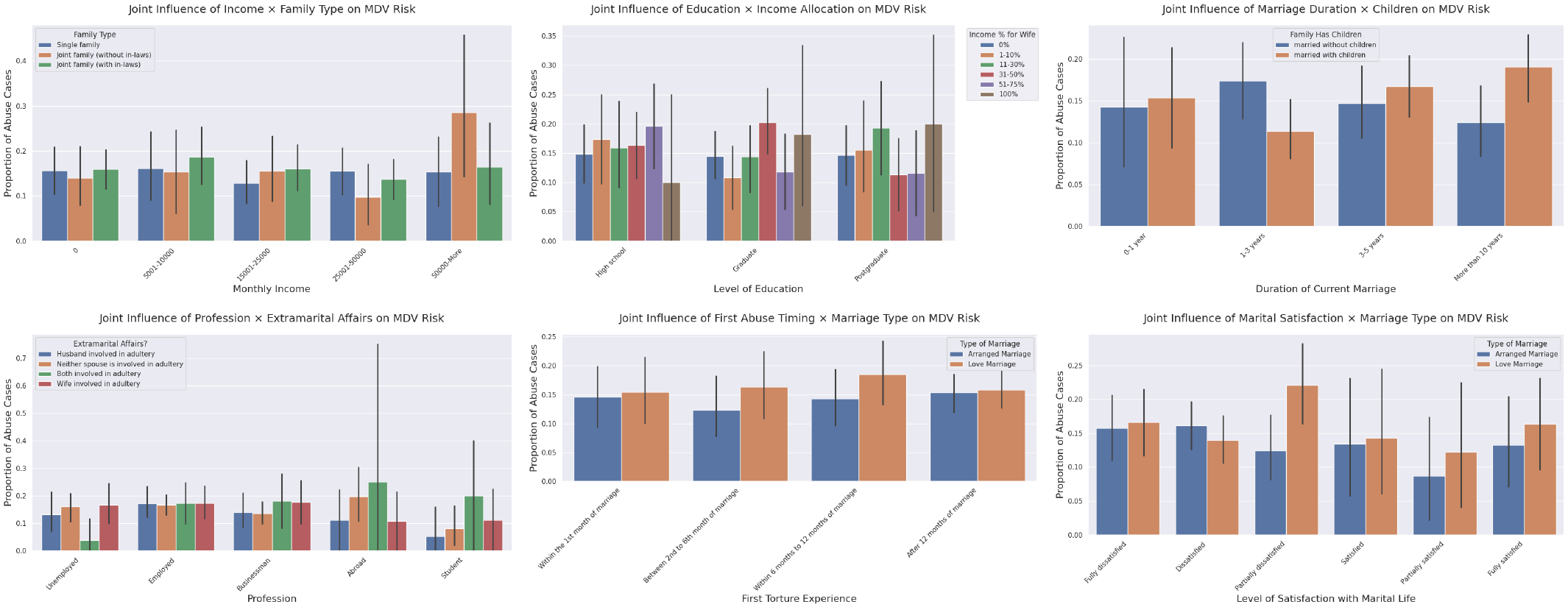}
    \caption{EDA of pairwise feature interactions. The plots show the proportion of abuse cases across different combinations of socio-demographic and relational factors, revealing complex, non-linear relationships that justify the need for an advanced ML approach.}
    \label{fig:eda_interactions}
\end{figure*}

To move beyond single-variable distributions, we further explored the interplay between key features by visualizing their pairwise interactions in Fig.~\ref{fig:eda_interactions}. These plots reveal that the influence of a single factor is often highly contingent on another. For example, the chart for monthly income versus family type (upper left) shows a complex relationship where the risk of abuse for those in the highest income bracket (50000-more) is notably higher in joint families. The plot for marital satisfaction versus marriage type (lower right) illustrates a critical interaction: while marital dissatisfaction is a general risk factor, its impact is far more pronounced in love marriages, which exhibit the highest proportion of abuse cases in the Partially dissatisfied category. Furthermore, the plot for marriage duration versus the presence of children (upper right) suggests that children have a non-linear effect, appearing to reduce risk in the early years of a marriage (1-3 years) but being associated with higher risk in the longest marriages (More than 10 years). Collectively, these visual analyses underscore the complex, combinatorial nature of MDV, highlighting the limitations of single-variable analysis and providing a strong rationale for applying a sophisticated machine learning model capable of capturing these intricate patterns.

\subsubsection{Cramér’s V Correlation}
%10
\begin{figure*}[!ht]
    \centering
    \includegraphics[width=1\linewidth]{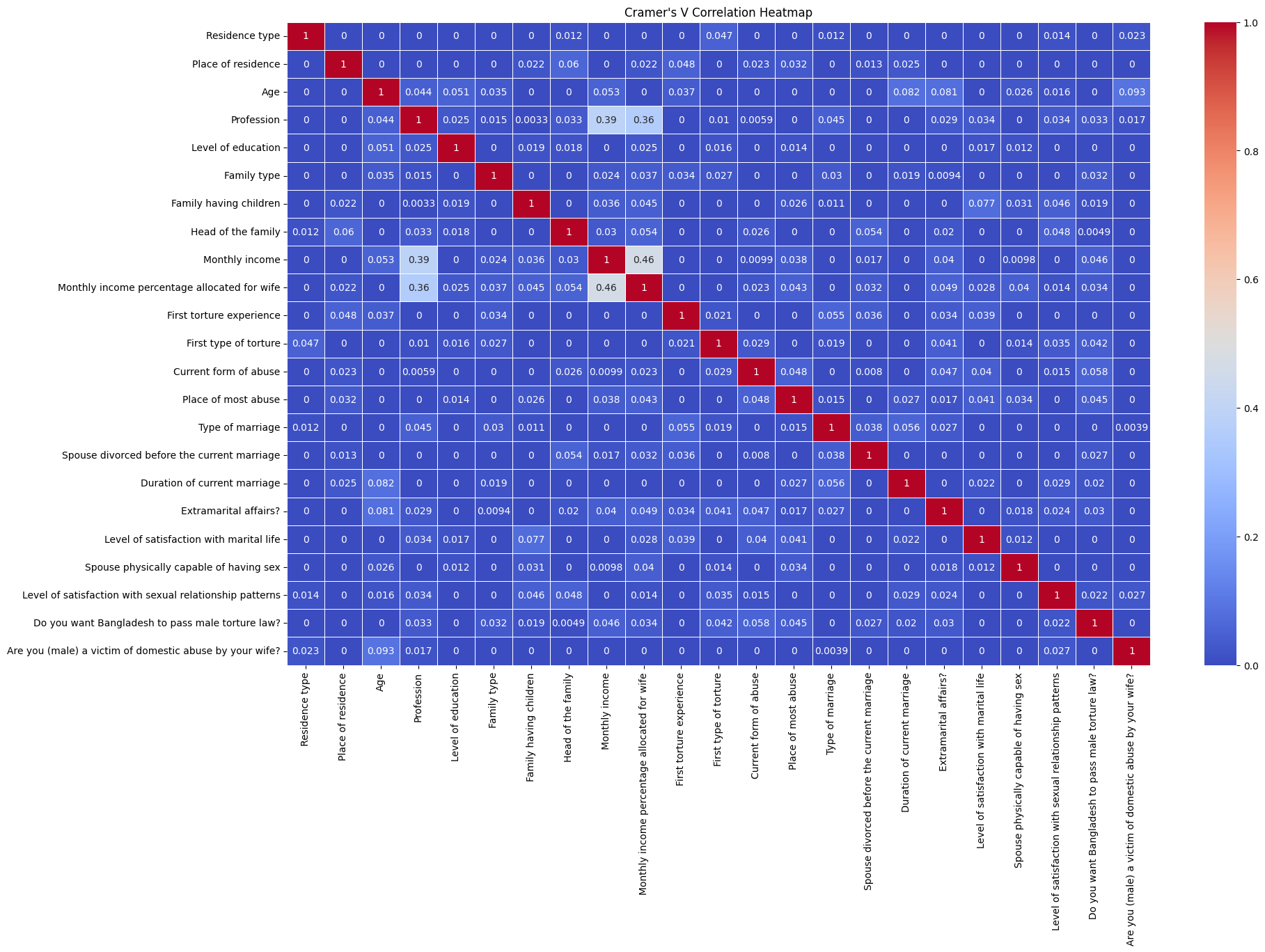}
    \caption{Heatmap of Cramer's V correlation matrix.}
    \label{fig:7}
\end{figure*}

Fig.~\ref{fig:7} presents the Cramér’s V correlation heatmap, which provides quantitative insights into the relationships between categorical variables in the dataset, with most correlations being weak ($<0.1$) but a few showing moderate associations. Notably, the correlation between ``Monthly income" and ``Monthly income percentage allocated for wife" (0.46) highlights the influence of financial capacity on resource allocation within households. The correlation between ``Profession" and ``Monthly income" (0.39) indicates that professional status significantly impacts income levels, reflecting the economic stratification based on employment types. Meanwhile, the correlation between ``Monthly income percentage allocated for wife" and ``Profession" (0.36) emphasizes how household income allocation may also depend on the earning member’s professional standing. While these correlations indicate significant relationships between the variables, it's important to note that Cramér’s V measures association strength between categorical variables and does not imply causation. Therefore, further analysis is required to understand the underlying mechanisms that drive these correlations. Correlations close to zero indicate minimal linear relationships between corresponding variables. This suggests that these variables may not have a strong association with each other based on Cramér’s V correlation coefficient.

\subsubsection{Chi-Square Test for Association Analysis}
In this subsection, we employ the Chi-square ($\chi^2$) test to explore potential associations between various categorical variables and the target variable, which identifies male victims of domestic abuse by their wives. The rationale behind this test was to investigate whether certain demographic, socioeconomic, or relational factors were significantly associated with male victimization in domestic abuse.

We conducted the $\chi^2$ test with a predetermined significance level ($\alpha$) of 10\%, as shown in Table \ref{chi-square}. This significance level was chosen to confirm a relatively liberal threshold for identifying potential associations while maintaining statistical rigor. We computed the $\chi^2$ statistic and corresponding p-value for each categorical variable to assess the likelihood of observing the data under the null hypothesis of independence between the variable and the target. Variables with p-values lower than the significance level were deemed to show a significant association with the target. In comparison, those with higher p-values did not have a significant association.

\begin{table*}[!ht]
\centering
\caption{$\chi^2$ test results for the association between the target variable and the other features.}
\label{chi-square}
\resizebox{\linewidth}{!}{%
\begin{tabular}{llll}
\toprule
\textbf{Feature} & \textbf{$\chi^2$ value} & \textbf{p-value} & \textbf{Decision} \\  \midrule
Residence type& 3.019660992& 0.220947426& No significant association
\\
Place of residence& 7.487215294& 0.485089517& No significant association
\\
Age& 42.21349485& 0.017045964& Significant association
\\
Profession& 4.609167901& 0.329798516& No significant association
\\
Level of education& 0.601231918& 0.740362047& No significant association
\\
Family type& 0.311298677& 0.855859254& No significant association
\\
Family having children& 0.267159624& 0.605243516& No significant association
\\
Head of the family& 0.119513024& 0.729563255& No significant association
\\
Monthly income& 3.31293946& 0.50688473& No significant association
\\
Monthly income percentage allocated for wife& 1.658140038& 0.894130624& No significant association
\\
First torture experience& 0.51854236& 0.914799255& No significant association
\\
First type of torture& 2.425301266& 0.787700402& No significant association
\\
Current form of abuse& 4.155071592& 0.527312484& No significant association
\\
Place of most abuse& 0.292503494& 0.863940181& No significant association
\\
Type of marriage& 1.0308213& 0.309965846& No significant association
\\
Spouse divorced before the current marriage& 1.885914695& 0.596419695& No significant association
\\
Duration of current marriage& 1.772013287& 0.621044012& No significant association
\\
Extramarital affairs?& 0.639233603& 0.887394877& No significant association
\\
Level of satisfaction with marital life& 3.104939202& 0.683811138& No significant association
\\
Spouse physically capable of having sex& 1.101579376& 0.77669283& No significant association
\\
Level of satisfaction with sexual relationship patterns& 6.468254567& 0.263282648& No significant association
\\
Do you want Bangladesh to pass a male torture law?& 0.009645347& 0.921764962& No significant association
\\ \bottomrule
\end{tabular}
}
\end{table*}

Upon conducting a $\chi^2$ test, we observed several interesting findings. Most analyzed variables did not show a statistically significant direct association with the target variable. Contrary to some existing literature on DV in other contexts, key factors such as `Profession,' `Level of education,' and `Level of satisfaction with marital life' did not yield significant p-values. The only variable exhibiting a simple significant association was `Age' ($p = 0.017$). While some variables like Residency type ($p = 0.2209$) and Level of satisfaction with sexual relationship patterns ($p = 0.2633$) had relatively lower p-values, they did not reach statistical significance. Other features, including First torture experience and Type of marriage, showed no strong associations with the target variable. 

This lack of significant findings for otherwise important variables is, in itself, a critical insight. It strongly suggests that the influence of these factors on MDV in the Bangladeshi context is not straightforward or linear. Rather than having a strong, independent predictive power, their impact is likely contingent on complex interactions with other variables. For instance, the effect of education may only become apparent when considered alongside income, family type, and place of residence. This outcome shows the limitations of traditional bivariate statistical tests for capturing the multifaceted nature of MDV. It thereby provides a compelling rationale for the application of advanced ML models, as employed in this study, which are specifically designed to uncover such complex, multivariate, and non-linear patterns that simpler tests would miss.

\subsection{Data Preprocessing}
The data preprocessing phase in our investigation was essential to confirm the readiness of the dataset for analysis. Categorical features within the dataset were processed to transform them into a numerical format compatible with ML algorithms. Using label Encoding, each categorical variable was converted into a numeric representation. This transformation facilitates the integration of categorical data into the modeling process without compromising the integrity of information. All data entries were converted to a consistent numerical format to maintain uniformity and compatibility across features. This step was crucial to confirm the seamless application of ML models and analytical techniques to the dataset. To handle the categorical features, including those with high cardinality, such as `Place of residence' and `Profession', we employed Label Encoding, where each unique category within a feature was assigned a unique integer. We chose this method over alternatives like one-hot encoding primarily to manage dimensionality. Given that our dataset contains several features with many unique categories, one-hot encoding would have drastically increased the feature space, potentially leading to sparsity and increased computational cost. Since our primary models, particularly tree-based ensembles like CatBoost and Random Forest, can effectively handle integer-encoded categorical variables without assuming an ordinal relationship, Label Encoding provided an efficient and effective solution for our modeling framework. 

Then, the dataset was divided into training and test sets, with an initial split of 80\% for training and 20\% for testing. The original class distribution in the dataset showed a significant imbalance, with 1692 non-abused instances (label 0) and 308 abused instances (label 1), resulting in a ratio of approximately 5:1 (Non-Abused: Abused). To address this imbalance, the training data was resampled using the Synthetic Minority Over-sampling Technique (SMOTE), which increased the number of abused instances to match the number of non-abused instances, creating a balanced 1:1 ratio. Following resampling, the training dataset consisted of 1692 non-abused and 1692 abused instances. Similarly, the test dataset, after resampling, contained 338 non-abused and 339 abused instances, also achieving a balanced class distribution. To confirm a robust and unbiased evaluation and to guard against overfitting, a strict validation protocol was followed. The 20\% test set was held out as a final, unseen dataset, completely isolated from all training, tuning, and resampling procedures. To prevent data leakage from the oversampling process, SMOTE was integrated into a \texttt{scikit-learn} pipeline within our 10-fold cross-validation loop. This confirms that for each validation fold, oversampling was applied \textit{only} to the training partition of that fold, while the validation partition remained in its original, imbalanced state. This best-practice approach guarantees that the model is always tuned and validated against data that reflects the original, real-world class distribution, providing a reliable estimate of its generalizability.

\subsection{Model Development}
In this study, we thoroughly examined various ML models, which involved a systematic evaluation process that included traditional ML classifiers, DL models, and ensemble models.

\subsubsection{ML Models}
In the initial phase of our analysis, we evaluated the performance of traditional ML classifiers with their default hyperparameters, establishing baseline performance to understand their predictive capabilities. Subsequently, we fine-tuned these classifiers using RandomizedSearchCV to optimize their hyperparameters. This technique involves systematically exploring defined hyperparameter spaces by randomly sampling combinations and selecting the best based on the chosen performance metric, which is accuracy in our case. We predefined a range of crucial hyperparameters for each classifier, such as regularization strength, tree depth, and learning rate. RandomizedSearchCV then conducted an iterative search, evaluating each combination using 10-fold cross-validation. After completing 10 iterations, the algorithm identified the optimal hyperparameters, which were then applied to the classifiers to improve their performance. The best hyperparameters for each classifier are summarized in Table~\ref{tab:best_hyperparameters}.

\begin{table*}[!ht]
\centering
\caption{Best hyperparameters selected by RandomizedSearchCV}
\label{tab:best_hyperparameters}
\resizebox{\linewidth}{!}{%
\begin{tabular}{l l}
\hline
\textbf{Model} & \textbf{Best hyperparameters} \\ \hline
LR & penalty = l1 ,  C=0.1\\ 
DT & min\_samples\_split = 10 , max\_depth = 10\\ 
RF & n\_estimators = 500 , max\_depth = 20\\ 
GBM & n\_estimators = 200 , learning\_rate=  0.01\\ 
SVM & kernel=  `rbf', C=  10\\ 
NB & priors = None, var\_smoothing = 1e-09 \\ 
KNN & weights = `distance', n\_neighbors =  3\\ 
ANN & hidden\_layer\_sizes = (512, 256, 128), activation = `tanh'\\ 
XGBoost & subsample = 0.8, n\_estimators = 200, max\_depth = 10, learning\_rate = 0.1, colsample\_bytree = 0.8\\ 
LGBM & subsample = 0.8, n\_estimators = 500, max\_depth = 10, learning\_rate = 0.01, colsample\_bytree = 0.8\\ 
Catboost & learning\_rate = 0.1 , l2\_leaf\_reg = 1, iterations = 500, depth = 10, bagging\_temperature = 1.0\\ \hline
\end{tabular}
}
\end{table*}

\paragraph{Logistic Regression (LR)}
LR is a statistical technique commonly used for binary classification tasks. It employs the sigmoid function, an S-shaped curve, to transform the input values into probabilities between 0 and 1. This function is the foundation of the LR model, which predicts the probability of an event with two possible outcomes. The following equation represents the LR model:
\begin{align}
P(y=1) &= \frac{1}{1 + e^{-(b_0 + b_1x)}}
\end{align}
where \( P(y=1) \) represents the probability of a positive outcome, \( b_0 \) is the intercept, \( b_1 \) is the coefficient associated with predictor variable \( x \), and \( e \) is the base of the natural logarithm. In contrast, the linear model used in ordinary least squares regression is denoted as:
\begin{align}
y &= b_0 + b_1x
\end{align}
where \( y \) represents the dependent variable, \( b_0 \) is the intercept, \( b_1 \) is the coefficient associated with predictor variable \( x \), and \( x \) is the independent variable.

\paragraph{Decision Tree (DT)}
DT is a non-parametric supervised learning approach commonly used for classification and regression tasks. It is structured hierarchically and comprises root nodes, branches, internal nodes, and leaf nodes. Each node in the tree represents an event or decision point, whereas the edges connecting the nodes depict the decision rules or conditions. The algorithm aims to minimize impurities from the root node to the leaf nodes to construct an effective DT. This is typically achieved using impurity measures such as the Gini index or Information Gain (IG). The Gini Index measures the probability of incorrectly classifying a randomly chosen element if it is labeled according to the distribution of labels in the node. Conversely, IG quantifies the reduction in entropy or uncertainty achieved by splitting data based on a particular attribute. When training a DT, the goal is to find the optimal split at each node by maximizing IG. IG is calculated using the following formula:
\begin{align}
    E(s) = \sum_{i=1}^{c} -p_i \log_2(p_i)
\end{align}
where, \( p_i \) represents the probability of class \( i \) in the dataset. By calculating the entropy of the dataset before and after a split and subtracting the two, we obtained the IG associated with that split. Maximizing the IG results in more effective splits and a more accurate DT model.

\paragraph{Random Forests (RF)}
RF is a powerful ensemble learning method that integrates the results of multiple DTs into a single outcome. In classification tasks, RF employs a majority voting mechanism, whereas in regression tasks, it calculates the average prediction of all individual trees. The RF trains each DT using the bagging method, in which each tree is trained on a random subset of the training data with replacement. This approach improves the robustness and generalization ability of the model, making it suitable for various data types and complexities. Mathematically, an RF model can be represented as:
\begin{align}
\text{rf} = \sum_{i=1}^{n} T_i
\end{align}
where, \( T_i \) is the \( i^{th} \) DT in the forest. Each DT \( T_i \) contributed to the overall prediction of the RF model, and the final prediction was determined by aggregating the predictions of all individual trees.

\paragraph{Gradient Boosting (GBM)}
GBM method operates by iteratively selecting a function that moves toward a weak or negative gradient, aiming to minimize a given loss function. This iterative process incorporates the concept of a learning rate, gradually reducing each iteration's error rate. Mathematically, the probability \( p(c_1) \) of class 1 is calculated using the odds ratio:
\begin{align}
    p(c_1) = e^{\log(\text{odds}_1)} + e^{\log(\text{odds})}
\end{align}
where the odds ratio is determined by the ratio of instances belonging to class 1 to those belonging to class 2, which encapsulates the probability of class membership based on the odds of each class, facilitating the classification process within the GBM framework.

\paragraph{SVM}
SVMs leverage the kernel trick to perform non-linear classification tasks proficiently by implicitly projecting input data into high-dimensional feature spaces. This technique allows SVMs to draw the boundaries between classes in complex datasets. The margins separating these classes were strategically placed to minimize classification errors while maintaining the shortest possible distance to the data points. Mathematically, SVMs aim to minimize the following objective function:
\begin{align}
\min_{w, b, \xi} \frac{1}{2} \|w\|^2 + C \sum_{i=1}^{n} \xi_i
\end{align}
where \(w\) represents the weight vector, \(b\) is the bias term, \(\xi\) denotes the slack variables, and \(C\) is a hyperparameter that controls the trade-off between maximizing the margin and minimizing the classification error. The term \(\frac{1}{2} \|w\|^2\) penalizes the magnitude of the weight vector \(w\), encouraging simpler decision boundaries, whereas \(C \sum_{i=1}^{n} \xi_i\) penalizes misclassified data points (\(\xi_i\)). By optimizing this objective function, SVMs effectively learn decision boundaries that separate classes while maximizing the margin between them, improving their classification accuracy on linear and non-linear datasets.

\paragraph{Naive Bayes (NB)}
NB  is a classification method that uses the Bayes theorem and the independent predictor premise. This classifier assumes that the existence of one feature in a class is unrelated to the presence of any other feature. It is mostly used for classification and clustering purposes based on the conditional likelihood of occurrence.
\begin{align}
    p(c|x) = P(x|c) P(c)P(x)
\end{align} 	                                
The posterior probability $P(c|x)$ is calculated from the Likelihood $P(x|c)$, class prior probability $P(c)$, and the predictor prior probability $P(x)$.

\paragraph{K-Nearest Neighbor (KNN)}
KNN  is a widely used supervised ML algorithm for classification and regression tasks. Its simplicity and intuitiveness make it attractive for various applications. Yet, a significant drawback of the KNN is its computational inefficiency, particularly as the volume of data increases. The core principle of the KNN involves determining the similarity between data points based on their features. This is typically achieved using distance metrics, with the Euclidean distance being a common choice. The Euclidean distance function used in the KNN is defined as:  
\begin{align}
\sqrt{\sum_{i=1}^{k} (x_i - y_i)^2}
\end{align}
where $x_i$ and $y_i$ represent the \( i^{th} \) features of the two data points, and $k$ is the number of features. This formula calculates the straight-line distance between two points in a multidimensional space. Although KNN's simplicity and effectiveness make it a popular choice, its performance can degrade significantly with large datasets because of the need to compute distances between the target point and all other data points. So, techniques such as dimensionality reduction or approximate nearest neighbor methods are often employed to mitigate this computational burden.

\paragraph{XGBoost}
Extreme GBM (XGBoost) is a popular implementation of GBMs. XGBoost is particularly favored because of its high execution speed and scalability. It works by sequentially adding DTs to an ensemble, with each tree aimed at correcting the errors of its predecessors. The final prediction was the weighted sum of the predictions from all the trees in the ensemble. The mathematical formulation of XGBoost is as follows:
\begin{align}
A_i = \phi(x_i) = \sum_{k=1}^{K} f_k(x_i)
\end{align}
Here, for each instance $x_i$ in the dataset ($x_i$, $y_i$), $y_i$ represents the predicted output. The term $f_k$ denotes the prediction of the $k^{th}$ tree, and $K$ is the total number of trees in the ensemble. The function $\phi(x_i)$ represents the final prediction, for instance $x_i$, which is the sum of predictions from all trees. XGBoost aims to determine the optimal collection of functions $f_k$ by minimizing a predefined loss function and regularization term.

\paragraph{LightGBM (LGBM)}
The LGBM is another implementation of GBM techniques. It differs from XGBoost and traditional GBM, using a histogram-based approach for tree construction. Instead of continuously binning the data, LGBM bins the data based on the histogram of the feature values. This histogram-based approach accelerates the training process by reducing the computational cost of finding the best split points. LGBM is well-suited for sparse datasets because it efficiently handles missing values and reduces memory usage.

\paragraph{CatBoost}
CatBoost is an algorithm based on DTs and GBM. It excels in handling categorical features efficiently and effectively. CatBoost improves the algorithm's generalization ability and robustness by mitigating gradient bias and prediction shift concerns. This is accomplished by incorporating the concept of GBM, which involves iteratively fitting new models to the residual errors generated by previous models. Mathematically, the objective function of CatBoost can be represented as:
\begin{align}
    \text{Obj}(\mathbf{w}) = \sum_{i=1}^{n} L(y_i, F(\mathbf{x}_i)) + \sum_{i=1}^{k} \Omega(f_i) + \sum_{i=1}^{m} \gamma Q_i
\end{align}
where \( \text{Obj}(\mathbf{w}) \) is the objective function to be minimized, \( L(y_i, F(\mathbf{x}_i)) \) represents the loss function that measures the discrepancy between the predicted values \( F(\mathbf{x}_i) \) and the actual labels \( y_i \), \( \Omega(f_i) \) is the regularization term penalizing the complexity of individual trees \( f_i \), \( \gamma \) is the regularization parameter, \( Q_i \) denotes the splitting criterion for each node in the trees.

During the learning process, CatBoost employs a technique known as ``Boost" to analyze categorical features. This involves considering all the sample datasets and randomly permuting them to create subsets. Subsequently, it filters out samples for each attribute belonging to the same group. Mathematically, this process can be represented as:
\begin{align}
    \text{Boost}(\mathbf{x}_i, g) = \text{Permute}(\text{Group}(\mathbf{x}_i, g))
\end{align}
where \( \text{Boost}(\mathbf{x}_i, g) \) represents the boost operation on the input feature \( \mathbf{x}_i \) within group \( g \), \( \text{Permute}(\cdot) \) denotes the permutation operation, \( \text{Group}(\mathbf{x}_i, g) \) partitions the samples of feature \( \mathbf{x}_i \) into groups based on categorical values. CatBoost can efficiently handle categorical data by incorporating Boost, enabling rapid processing and robust model performance.

\subsubsection{DL and Ensemble Models}
We explored the efficacy of DL and ensemble techniques to maximize predictive accuracy, explicitly focusing on stacking classifiers. Stacking classifiers offers a sophisticated approach to combining the strengths of multiple base classifiers to achieve superior predictive performance. We evaluated two distinct stacking ensembles, each tailored to leverage the unique strengths of its constituent models. The base classifiers for stacking were selected based on their area under the curve (AUC) scores, ensuring that only the most effective models contributed to the ensemble.

\paragraph{Artificial Neural Networks (ANN)}
ANNs are inspired by the structure and functionality of the human brain. These models consist of interconnected nodes, or neurons, organized into layers, including input, hidden, and output layers. Each neuron in the network applies an activation function to its inputs, allowing the ANN to capture and model complex patterns in the data. Our study employed the MLPClassifier from the Scikit-learn library to implement an ANN. The model was configured with two hidden layers comprising 512 and 256 neurons, respectively, and the tanh activation function was used to introduce non-linearity. The network was trained for a maximum of 500 iterations, enabling it to adapt and generalize to a variety of input patterns effectively.

\paragraph{Fully Connected Network (FCN)} The FCN model is implemented using the TensorFlow library and is structured as a feedforward neural network. The architecture comprises three dense layers: a 512-neuron input layer, a 256-neuron hidden layer, and a single output neuron for binary classification. The activation function used in the input and hidden layers is the rectified linear unit (ReLU), introducing non-linearity to allow the model to learn complex data patterns. The output layer employs a sigmoid activation function, which outputs a probability score for the positive class. The model is compiled using the binary cross-entropy loss function, the Adam optimizer, and binary accuracy as the evaluation metric.

\paragraph{BiLSTM+FCN} This hybrid DL model integrates a BiLSTM with dense FCN to process sequential data effectively. The input sequences, preprocessed with label encoding for categorical features and padded to a fixed length of 64, are passed through an embedding layer with 128-dimensional dense vector representations. Two BiLSTM layers, with hidden units of 64 and 512, respectively, capture both forward and backward temporal dependencies. Dense layers with 64, 32, and 16 neurons, activated by ReLU functions, extract hierarchical feature representations. A dropout layer with a rate of 0.2 prevents overfitting, and the final classification layer, activated by softmax, outputs probabilities for two classes. The model is compiled with the Adam optimizer (learning rate = 0.01) and trained using categorical cross-entropy loss and categorical accuracy as the evaluation metric. To address data imbalance, class weights were calculated as 5:1 for the non-abused and abused classes. The training was performed with a batch size of 32 and an early stopping callback, which monitored validation loss with patience of 5 epochs and restored the best weights. The model effectively balances performance and efficiency while addressing the challenges posed by class imbalance and overfitting.

\paragraph{ANN-CatBoost-RF Stacking + LR} This model integrates ANN, CatBoost, and RF as base classifiers, with LR serving as the meta-classifier. These three models--ANN, CatBoost, and RF--were chosen as the base classifiers based on their superior individual performance according to the AUC curve. ANN excels at identifying complex, non-linear relationships in the data. CatBoost is particularly adept at handling categorical features and imbalanced datasets, and RF offers robustness against overfitting. The predictions from these base models are aggregated and synthesized by LR, which combines their outputs into a final prediction. This ensemble approach improves predictive accuracy by leveraging the complementary strengths of the base classifiers, as validated through comprehensive performance metrics, including AUC and confusion matrices.

\paragraph{ANN-CatBoost Stacking + LR} This model simplifies the stacking structure by focusing on ANN and CatBoost as the base classifiers, also employing LR as the meta-classifier. This reduced model prioritizes computational efficiency while maintaining competitive performance. The model minimizes redundancy by selecting ANN and CatBoost, which consistently achieve the best AUC scores individually (see Table~\ref{tab:performance}) and confirm the ensemble captures the most significant predictive signals. The streamlined architecture demonstrated strong classification metrics, making it a resource-efficient alternative for predictive tasks.

\subsection{Interpretability Analysis}
Our research incorporates XAI methods to improve the clarity and understanding of the proposed ensemble model, which integrates an ANN and CatBoost in a stacking architecture with LR as the meta-classifier. XAI focuses on enabling AI systems to provide clear reasons for their decisions and predictions, which is essential for building user trust and comprehension. 

\subsubsection{Post-hoc Interpretability Using SHAP and LIME}
We employed two post-hoc XAI methods, LIME and SHAP, which allowed local and global interpretability of our stacking model's predictions, addressing the ``black-box" nature of our proposed ensemble model.

\paragraph{LIME}
LIME provides interpretability by approximating the predictions of a complex model with an interpretable surrogate model for a single instance. For our stacking model, the \textit{LimeTabularExplainer} was instantiated using the training dataset to build a locally interpretable linear model around each instance. 

Let \( \hat{y}_i = f(x_i) \) represent the prediction for an instance \( x_i \) made by the stacking model \( f \). LIME perturbs the features of \( x_i \) to generate a set of synthetic instances \( S \) and the corresponding predictions \( f(S) \). Using these, it fits a local surrogate model \( g \), typically a simple linear regression, around \( x_i \). The weights \( w(z_i) \) in \( g \) are derived from a kernel function \( \pi(z_i, x_i) \), where \( z_i \) represents a perturbed instance:
\begin{equation}
\pi(z_i, x_i) = \exp \left( - \frac{\| z_i - x_i \|^2}{\sigma^2} \right)
\end{equation}
Here, \( \sigma \) controls the width of the neighborhood around \( x_i \). The coefficients of \( g \) indicate the contribution of each feature to the prediction for \( x_i \). This approach was applied to selected test instances to provide insights into feature contributions for individual predictions.

\paragraph{SHAP}
Based on cooperative game theory, SHAP explains model predictions by attributing contributions of individual features to the output. SHAP values are computed as Shapley values, which fairly distribute the prediction among the input features. For a prediction \( \hat{y}_i \), the SHAP value \( \phi_j \) for feature \( x_j \) is defined as:
\begin{equation}
\phi_j = \sum_{S \subseteq F \setminus \{j\}} \frac{|S|!(|F| - |S| - 1)!}{|F|!} \left[ f(S \cup \{j\}) - f(S) \right]
\end{equation}
Here, \( F \) is the set of all features, \( S \) is a subset of features excluding \( x_j \), and \( f(S) \) is the model prediction when only features in \( S \) are used. This calculation is computationally expensive for high-dimensional data; hence, we used the SHAP Explainer, which approximates Shapley values efficiently. For our stacking model, we defined a prediction function \( f(x) \) corresponding to the meta-classifier's probability output. The SHAP Explainer generated global explanations by summarizing feature contributions across multiple instances. Local explanations for individual cases were visualized using waterfall plots and decision plots.

\subsubsection{Model-Specific Feature Importance for Interpretability}
In addition to employing SHAP and LIME for post-hoc interpretability, we utilized the model-specific feature importance functionality of the top three well-performing models, RF, CatBoost, and LightGBM, to evaluate feature contributions. These models inherently provide feature importance scores, which are calculated during the training process and offer insights into the global relevance of features.

For RF, feature importance was computed based on the decrease in impurity. Specifically, for a feature \( x_i \), the importance is calculated as:
\begin{equation}
\text{FI}(x_i) = \sum_{t \in T} \Delta I_t(x_i)
\end{equation}
where \( T \) represents the set of all trees in the forest, and \( \Delta I_t(x_i) \) is the total decrease in impurity contributed by \( x_i \) in tree \( t \). This decrease is defined as:
\begin{equation}
\Delta I_t(x_i) = \sum_{s \in S_i} (I_{\text{parent}}(s)(s) - I_{\text{left}}(s) - I_{\text{right}}(s))
\end{equation}
Here, \( S_i \) is the set of all splits involving \( x_i \), and \( I_{\text{parent}}(s)(s), I_{\text{left}}(s), I_{\text{right}}(s) \) are the impurity measures (e.g., Gini or entropy) for the parent and child nodes at split \( s \). For LightGBM, we utilized both split count and gain-based importance metrics. The split count importance was determined as:
\begin{equation}
\text{FI}(x_i) = \sum_{t \in T} \text{count}(x_i, t)
\end{equation}
where \( \text{count}(x_i, t) \) is the number of times \( x_i \) was used for splitting in tree \( t \). The gain-based importance was calculated as:
\begin{equation}
\text{FI}(x_i)) = \sum_{t \in T} \sum_{s \in S_i} G_s
\end{equation}
where \( G_s \) represents the loss reduction (gain) because of the split \( s \) involving \( x_i \). For CatBoost, feature importance was computed as the average reduction in the loss function attributed to each feature across all boosting iterations:
\begin{equation}
\text{FI}(x_i)) = \sum_{j=1}^n \frac{\Delta L_j(x_i)}{n}
\end{equation}
where \( n \) is the total number of iterations, and \( \Delta L_j(x_i) \) represents the reduction in the loss function because of \( x_i \) at iteration \( j \).

We utilized model-specific feature importance methods to address the limitation of SHAP and LIME when applied to ensemble models. While effective at explaining model outputs, these post-hoc interpretability tools do not provide transparency at the level of individual models that constitute an ensemble. By leveraging the feature importance metrics from RF, CatBoost, and LightGBM, we obtained global insights into feature relevance, complementing the local interpretability offered by SHAP and LIME. The computation of these feature importance scores is inherently efficient, as they are integrated into the model training process.

\section{Results and Discussions}
\label{sec:Results and Discussion}
\subsection{Model Evaluation Metrics}
We relied on a set of fundamental evaluation metrics to identify the most suitable model for this dataset concerning MDV in Bangladesh. These metrics provide nuanced insights into each model's performance, considering the problem's intricacies.

\paragraph{Accuracy} An essential metric measuring the proportion of correct predictions out of the total predictions made. Although useful, accuracy might not be reliable in cases of class imbalances, where one class significantly outweighs the other.
\begin{equation}
Accuracy = \frac{TN + TP}{TN + TP + FN + FP} 
\end{equation}

\paragraph{Macro Average Precision (maP)}  
Macro average precision quantifies the accuracy of positive predictions by calculating the precision for each class independently and then averaging these values. It is particularly useful in assessing model performance across imbalanced datasets.  
\begin{equation}
maP = \frac{1}{C} \sum_{i=1}^{C} \frac{TP_i}{TP_i + FP_i}
\end{equation}

\paragraph{Macro Average Recall (maR)}  
Macro average recall measures the model's ability to identify all relevant instances by calculating the recall for each class and averaging these values. This metric is vital for evaluating the sensitivity of models across imbalanced datasets.  
\begin{equation}
maR = \frac{1}{C} \sum_{i=1}^{C} \frac{TP_i}{TP_i + FN_i}
\end{equation}

\paragraph{Macro Average F1-Score (maF1)}  
The macro average F1-Score computes the harmonic mean of maP and maR, offering a balanced performance measure across all classes.  
\begin{equation}
maF1 = \frac{2 \times maP \times maR}{maP + maR}
\end{equation}

\paragraph{Area Under the Curve (AUC)} AUC measures the model's ability to distinguish between positive and negative classes. It evaluates the trade-off between true positive and false positive rates at various thresholds, providing an aggregated performance measure. A higher AUC indicates better discrimination capability.

\paragraph{Specificity} Also known as the True Negative Rate, specificity measures the proportion of actual negatives correctly identified by the model. It highlights the model's ability to avoid false positives, which is crucial in scenarios where minimizing the misclassification of the negative class is essential.
\begin{equation} Specificity = \frac{TN}{TN + FP}
\end{equation}

\paragraph{Geometric Mean (G-Mean)} G-Mean evaluates the balance between sensitivity (recall) and specificity. It is particularly valuable for assessing model performance in imbalanced datasets by penalizing models that perform well on one class but poorly on the other.
\begin{equation} G\text{-}Mean = \sqrt{Recall \times Specificity}
\end{equation}

\paragraph{Index of Balanced Accuracy (IBA)} IBA extends the concept of balanced accuracy by incorporating a weighting factor that prioritizes either sensitivity or specificity, depending on the application. It is a robust metric for imbalanced datasets, offering flexibility to emphasize certain performance aspects.

These metrics allowed us to comprehensively evaluate the performance of each model and make informed decisions regarding its suitability for the task at hand.

\begin{figure*}[!ht]
\centering
\subfloat[Traditional ML models with default hyperparameter settings.]{
 \includegraphics[width=0.47\linewidth]{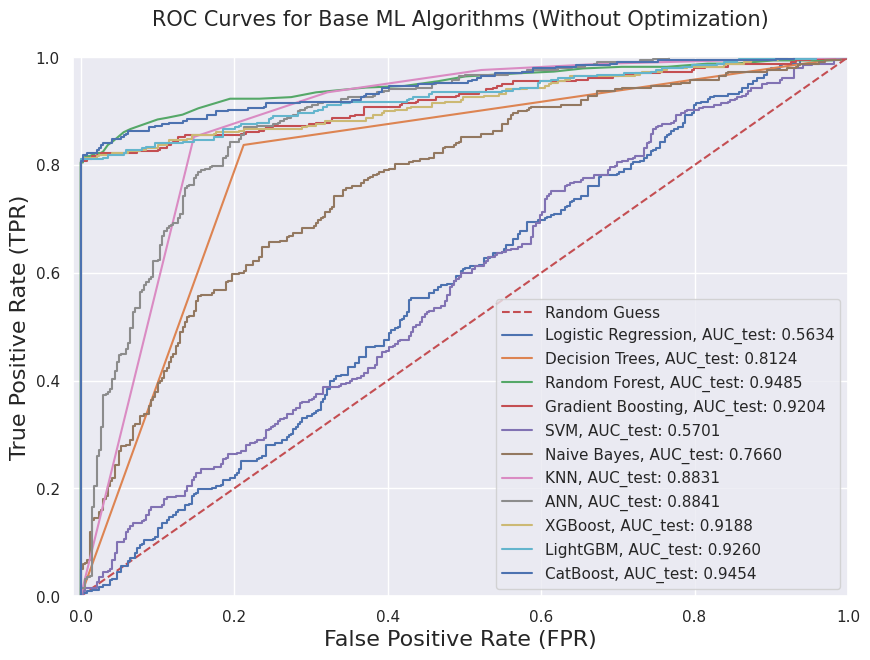}
 \label{fig:roc1}
}
\hfill
\subfloat[All the implemented models using the best hyperparameters.]{
 \includegraphics[width=0.47\linewidth]{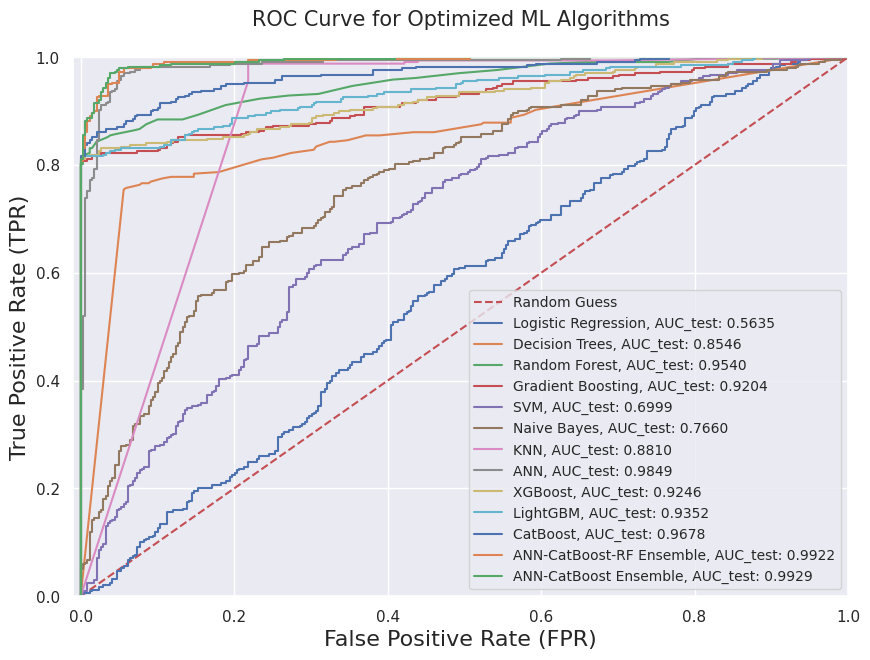}
 \label{fig:roc2}
}
\caption{ROC-AUC curves for (a) traditional ML models before hyperparameter optimization, and (b) traditional and hybrid models after hyperparameter optimization.}
\label{fig:roc-auc}
\end{figure*}

\subsection{Comparative Performance Analysis}
We evaluated the performance of 10 traditional ML models, 3 DL models, and 2 ensemble models for predicting MDV. Table~\ref{tab:performance} comprehensively evaluates traditional classifiers, DL models, and advanced ensemble techniques across various imbalanced classification metrics. Figure~\ref{fig:roc-auc} shows that after hyperparameter optimization using RandomizedSearchCV, the AUC improved in all the cases for the non-hybrid ML models. 

The traditional base classifiers demonstrated varying levels of effectiveness. LR achieved the lowest accuracy (55\%) and an AUC of 56.11\%, indicating its inability to effectively model non-linear relationships. On the other hand, DT, RF, and GB exhibited strong performance, with RF achieving the highest accuracy among the classical models (90\%) and an AUC of 95.61\%. Models like SVM and NB showed moderate results, with SVM achieving an accuracy of 64\% and an AUC of 69.99\%, and NB reaching 70\% accuracy with an AUC of 76.60\%. Meanwhile, KNN achieved 77\% accuracy but slightly improved AUC at 88.10\%. These findings show the robustness of tree-based models like RF and GB in capturing complex patterns and handling imbalanced datasets effectively.

DL models outperformed the traditional classifiers. ANN demonstrated exceptional accuracy (92\%) and a high AUC of 97.26\%. In the same way, FCN achieved 90\% accuracy with an AUC of 96.40\%. On the other hand, the hybrid DL model (BiLSTM+FCN) performed poorly, achieving only 50\% accuracy and an AUC of 64.69\%. This result highlights its limitations in effectively leveraging imbalanced datasets and extracting meaningful patterns.

The ensemble models significantly outperformed all others. The ANN-CatBoost-RF stacking model demonstrated an accuracy of 94\% with an AUC of 99.22\%, showcasing its superior ability to integrate the strengths of individual base models. However, the ANN+CatBoost stacking model with LR as the meta-model stood out as the best-performing approach. This model achieved the highest accuracy (95\%), precision, recall, F1-score (all at 95\%), and an exceptional AUC of 99.29\% (see Figure~\ref{fig:roc2}), indicating its unmatched ability to provide accurate and reliable predictions. Its performance in terms of specificity (98\%), G-Mean (94\%), and IBA (89\%) further highlights its robustness in minimizing false positives and negatives, as well as its ability to balance sensitivity and specificity.

The proposed ANN-CatBoost stacking + LR model combines the strengths of ANN's deep feature extraction with the interpretability and efficiency of CatBoost. By integrating these models through stacking and using LR as a meta-model, the proposed framework leverages the complementary strengths of its components. ANN effectively captures complex, non-linear relationships in the data, while CatBoost excels in handling categorical variables and imbalanced datasets. Including LR as a meta-model confirms the ensemble can optimally synthesize predictions, yielding state-of-the-art results.

\begin{table*}[!ht]
\centering
\caption{Performance metrics of traditional ML, DL, and ensemble models for predicting MDV. \textbf{Bold} indicates the best results for each metric, and \ul{Underline} indicates the second best results.}
\label{tab:performance}
\resizebox{\linewidth}{!}{%
\begin{tabular}{l c c c c c c c c} \hline          
\textbf{Model} & \textbf{Accuracy} & \textbf{maP} & \textbf{maR} & \textbf{maF1} & \textbf{AUC} & \textbf{Specificity} & \textbf{G-Mean} & \textbf{IBA} \\ \hline 
LR & 55\% & 55\% & 55\% & 54\% & 56.11\% & 52\% & 52\% & 52\% \\ 
DT & 82\% & 82\% & 82\% & 82\% & 85.60\% & 85\% & 81\% & 81\% \\ 
RF & 90\% & 92\% & 90\% & 90\% & 95.61\% & \textbf{100\%} & \ul{91\%} & \ul{90\%} \\ 
GBM & 89\% & 91\% & 89\% & 89\% & 91.79\% & \textbf{100\%} & 88\% & 88\% \\ 
SVM & 64\% & 65\% & 64\% & 64\% & 69.99\% & 53\% & 60\% & 59\% \\ 
NB & 70\% & 70\% & 70\% & 70\% & 76.60\% & 66\% & 69\% & 69\% \\ 
KNN & 77\% & 84\% & 77\% & 76\% & 88.10\% & 55\% & 70\% & 71\% \\ 
ANN & 92\% & 92\% & 92\% & 92\% & 97.26\% & 86\% & \ul{91\%} & \ul{90\%} \\ 
XGBoost & 90\% & 91\% & 90\% & 90\% & 92.46\% & \ul{99\%} & \ul{91\%} & \textbf{91\%} \\ 
LightGBM & 90\% & 91\% & 90\% & 90\% & 92.51\% & \ul{99\%} & \ul{91\%} & \textbf{91\%} \\ 
CatBoost & 91\% & 92\% & 91\% & 91\% & 96.78\% & \ul{99\%} & \ul{91\%} & \textbf{91\%} \\ 
FCN & 90\% & 91\% & 90\% & 90\% & 96.40\% & 90\% & 90\% & 81\% \\ 
BiLSTM+FCN & 50\% & 25\% & 50\% & 33\% & 64.69\% & 50\% & 0\% & 0\% \\ 
ANN-CatBoost-RF stacking + LR & \ul{94\%} & \ul{94\%} & \ul{94\%} & \ul{94\%} & \ul{99.22\%} & 98\% & \textbf{94\%} & 89\% \\ 
\textbf{ANN-CatBoost stacking + LR (Proposed)} & \textbf{95\%} & \textbf{95\%} & \textbf{95\%} & \textbf{95\%} & \textbf{99.29\%} & 98\% & \textbf{94\%} & 89\% \\ \hline
\end{tabular}
}
\end{table*}

\subsection{Statistical Analysis for Model Performance}
To evaluate the statistical significance of the differences in model performance, we conducted paired \( t \)-tests comparing each model's accuracy against the proposed model, ANN-CatBoost stacking + LR (see Table~\ref{tab:t-test}). The analysis was based on results from a 10-fold cross-validation procedure, ensuring a robust evaluation of performance variability across folds. The null hypothesis (\( H_0 \)) posited that there is no significant difference in accuracy between the models (\( \mu_{\text{difference}} = 0 \)), while the alternative hypothesis (\( H_1 \)) suggested a significant difference (\( \mu_{\text{difference}} \neq 0 \)). A significance level (\( \alpha \)) of 5\% was used, and Bonferroni correction was applied to account for multiple comparisons, resulting in an adjusted \( \alpha \) level of 0.0036.

\begin{table}[!ht]
\centering
\caption{Statistical paired t-test comparison of model accuracies with ANN-CatBoost stacking + LR using 10-fold cross-validation and Bonferroni-adjusted $\alpha$ = 0.0036.}
\label{tab:t-test}
\resizebox{\linewidth}{!}{%
\begin{tabular}{l c c c} \hline     
\textbf{Model} & \textbf{t-value (Accuracy)} & \textbf{p-value (Accuracy)} & \textbf{Cohen's \( d \) (Accuracy)} \\ \hline 
LR & 121.1051686 & 9.06E-16 & 38.29681691 \\  
DT & 80.25698535 & 3.66E-14 & 25.37948718 \\  
RF & 10.2632747 & 2.88E-06 & 3.24553243 \\  
GBM & 16.26647032 & 5.57E-08 & 5.143909572 \\  
SVM & 91.64586441 & 1.11E-14 & 28.98096697 \\  
NB & 150.0921306 & 1.32E-16 & 47.46329915 \\  
KNN & 114.5227586 & 1.50E-15 & 36.21527612 \\  
ANN & 20.22896814 & 8.21E-09 & 6.396961404 \\  
XGBoost & 37.65104053 & 3.26E-11 & 11.90630443 \\  
LightGBM & 23.62297162 & 2.08E-09 & 7.470239543 \\  
CatBoost & 26.08610852 & 8.63E-10 & 8.249151821 \\  
FCN & 44.42596868 & 7.41E-12 & 14.04872483 \\  
BiLSTM+FCN & 323.2059984 & 1.32E-19 & 102.2067108 \\  
ANN-CatBoost-RF stacking + LR & 4.208968922 & 0.002276276 & 1.330992839 \\ \hline
\end{tabular}
}
\end{table}

The analysis revealed significant differences in accuracy for all models compared to ANN-CatBoost stacking + LR. Each \( t \)-value and corresponding \( p \)-value were calculated, and effect sizes were quantified using Cohen's \( d \). For models such as LR, DT, and NB, the \( t \)-values and Cohen's \( d \) were exceptionally high, indicating large effect sizes and substantial deviations from the performance of ANN-CatBoost stacking + LR. In contrast, while showing significant differences, the ANN-CatBoost-RF stacking + LR model exhibited the smallest effect size (\( d = 1.3310 \)), suggesting that it was the closest in performance to the proposed model.

The BiLSTM+FCN model had the highest \( t \)-value (323.2060) and effect size (\( d = 102.2067 \)), reflecting the most substantial deviation from ANN-CatBoost stacking + LR. Other models, such as XGBoost, LightGBM, and CatBoost, also demonstrated significant but relatively moderate deviations, highlighting their strong baseline performance. The statistical tests confirm that the proposed ANN-CatBoost stacking + LR model significantly outperforms other baseline models in accuracy.

\subsection{Interpretability Results}
\subsubsection{Model-Specific Interpretations of Top Three Classifiers}
\begin{figure*}[!ht]
\centering
\subfloat[ANN feature importance.]{
 \includegraphics[width=0.47\linewidth]{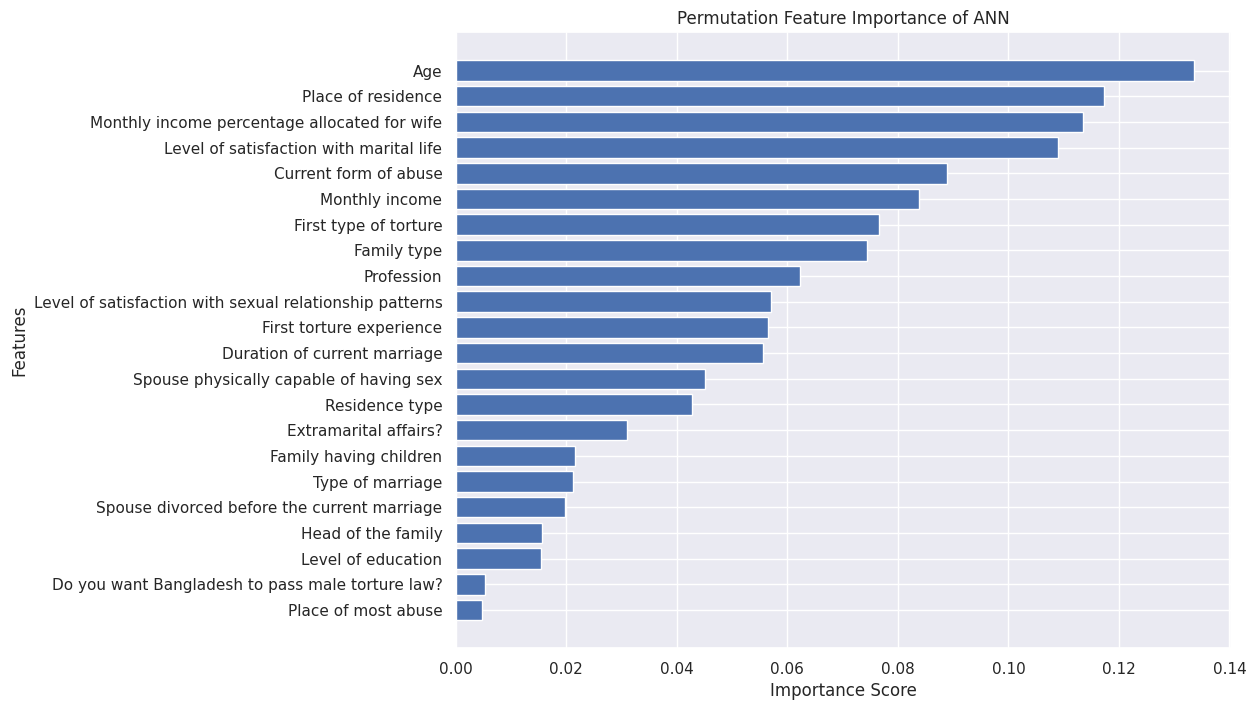}
 \label{fig:ann}
}
\hfill
\subfloat[CatBoost feature importance.]{
 \includegraphics[width=0.47\linewidth]{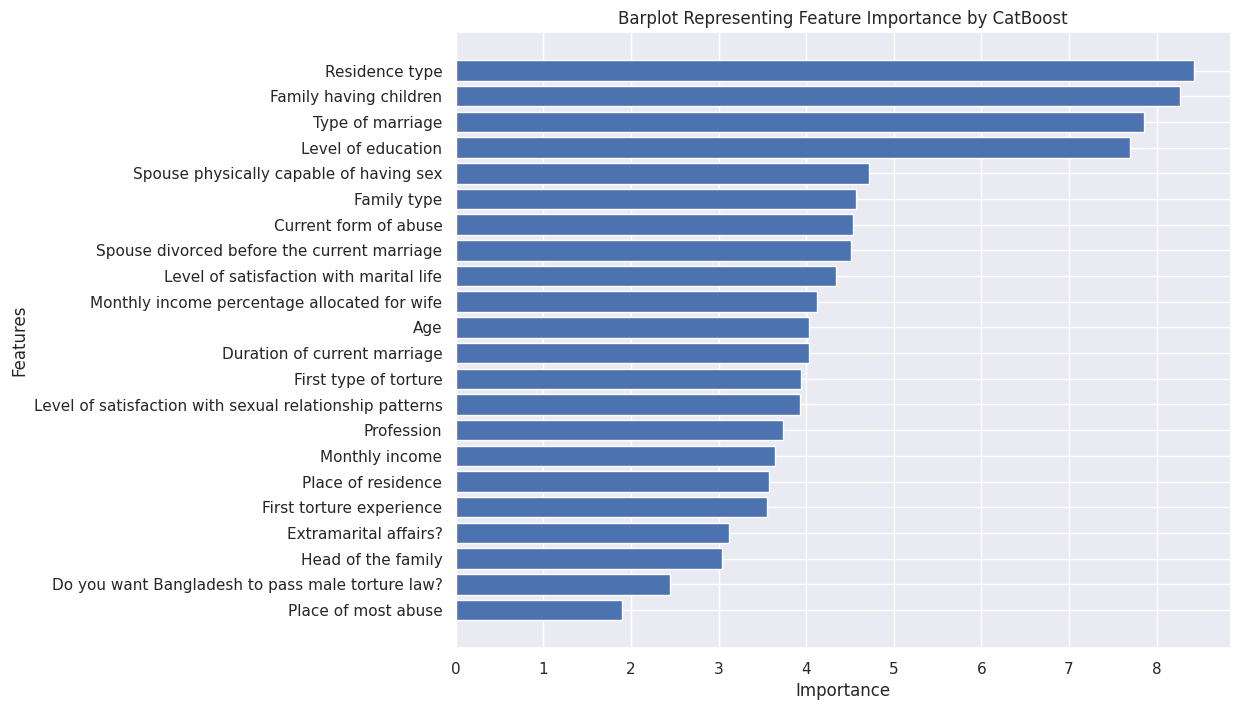}
 \label{fig:catboost}
}
\hfill
\subfloat[Random Forest feature importance.]{
 \includegraphics[width=0.47\linewidth]{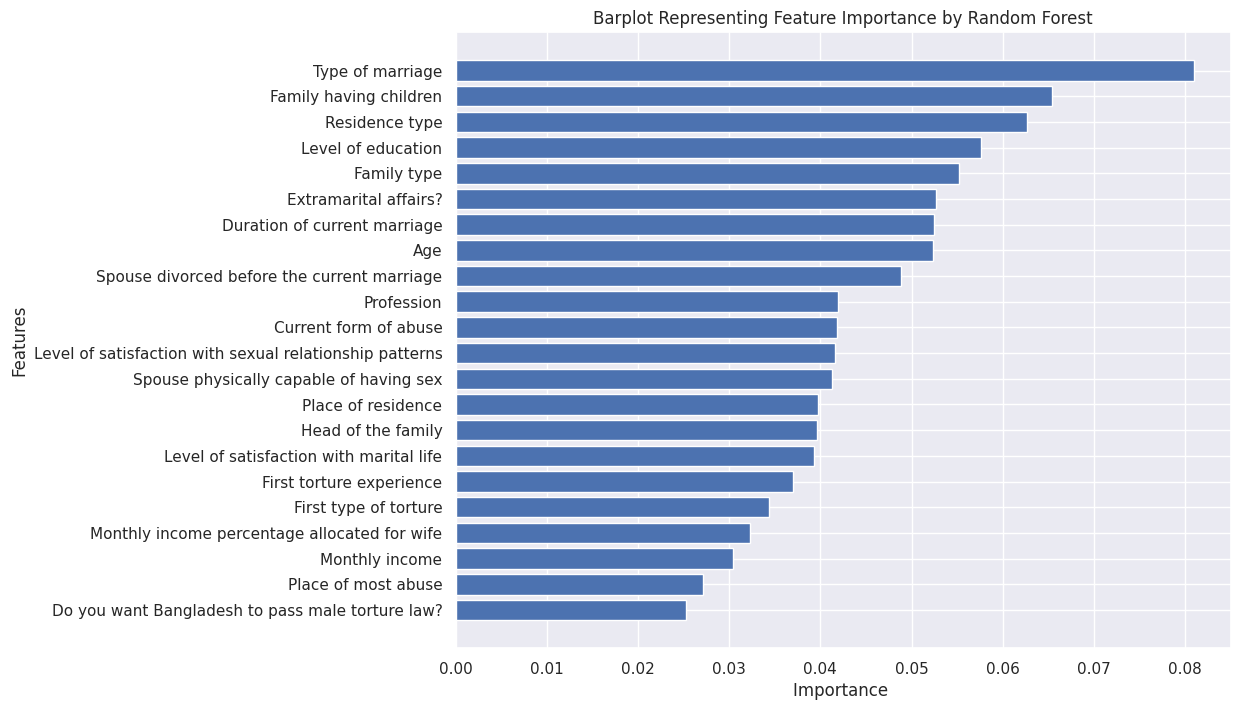}
 \label{fig:rf}
}
\caption{Feature importance analysis of the top three individual models selected based on their evaluation metrics.}
\label{fig:fi}
\end{figure*}

In Fig.~\ref{fig:fi}, we present three horizontal bar charts depicting the feature importance for three top-performing individual models: ANN, CatBoost, and RF. Each subfigure provides insights into the significance of individual features, with the x-axis representing importance scores and the y-axis listing the features.

Fig.~\ref{fig:ann} illustrates the feature importance derived from the ANN model, where factors such as age, place of residence, and monthly income percentage allocated for the wife emerge as the most influential features. These variables highlight the demographic and economic aspects that substantially impact the predictions. Fig.~\ref{fig:catboost} focuses on CatBoost, showcasing the prominence of features such as residence type, family having children, and type of marriage. This highlights the importance of living conditions and family dynamics in influencing the model's predictions. Fig.~\ref{fig:rf} depicts the feature importance of RF. Here, the type of marriage, the family having children, and the residence type stand out as key contributors. Other features, such as age and level of education, also hold notable significance, but to a lesser extent. The comparative analysis across these subfigures reveals variations in feature rankings, demonstrating how individual models prioritize features based on their inherent algorithmic structures and interactions, which is different from the ensemble model they constitute.

\subsubsection{SHAP Interpretations}
\begin{figure}[!ht]
    \centering
    \begin{subfigure}{0.49\textwidth}
        \centering
        \includegraphics[width=\textwidth]{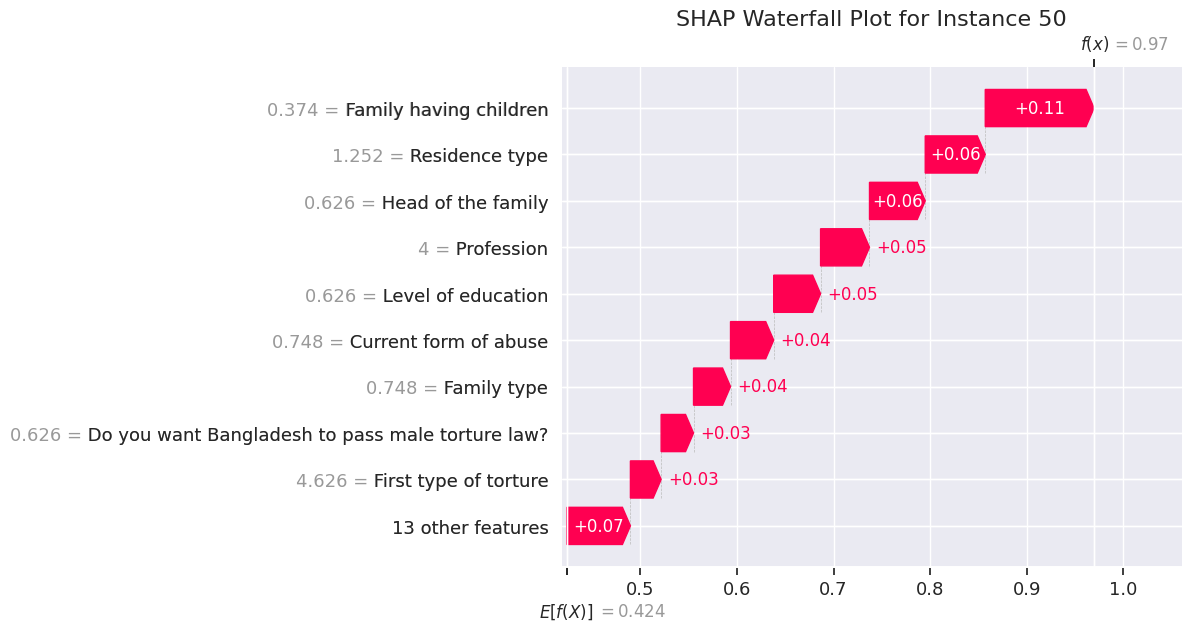}
        \caption{SHAP waterfall plot for instance 50.}
    \end{subfigure}
    \begin{subfigure}{0.49\textwidth}
        \centering
        \includegraphics[width=\textwidth]{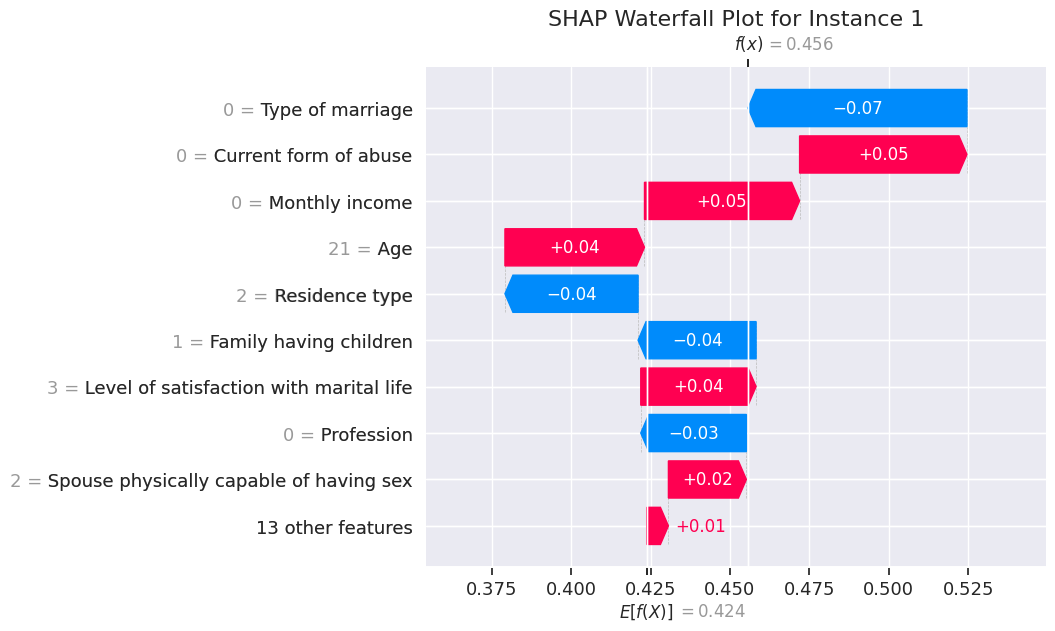}
        \caption{SHAP waterfall plot for instance 1.}
    \end{subfigure}
    \caption{Instance-level SHAP Waterfall Plots explaining predictions for specific instances.}
    \label{fig:instance_waterfall}
\end{figure}

To evaluate the explainability of the proposed model (ANN-CatBoost stacking + LR), SHAP was utilized to analyze feature contributions both at the instance level and across the entire dataset. Fig.~\ref{fig:instance_waterfall} displays two SHAP waterfall plots, which explain the contributions of individual features for specific instances. For `Instance 50', the plot demonstrates how features like \textit{Family having children}, \textit{Residence type}, and \textit{Profession} significantly contribute to the final prediction, resulting in a strong positive class prediction of 0.97. Likewise, for Instance 1, shown in the second waterfall plot, features such as \textit{Current form of abuse} and \textit{Monthly income} positively influence the prediction, while \textit{Type of marriage} has a negative contribution, leading to a final prediction score of 0.456.

\begin{figure*}[!ht]
    \centering
    \begin{subfigure}{0.49\textwidth}
        \centering
        \includegraphics[width=\textwidth]{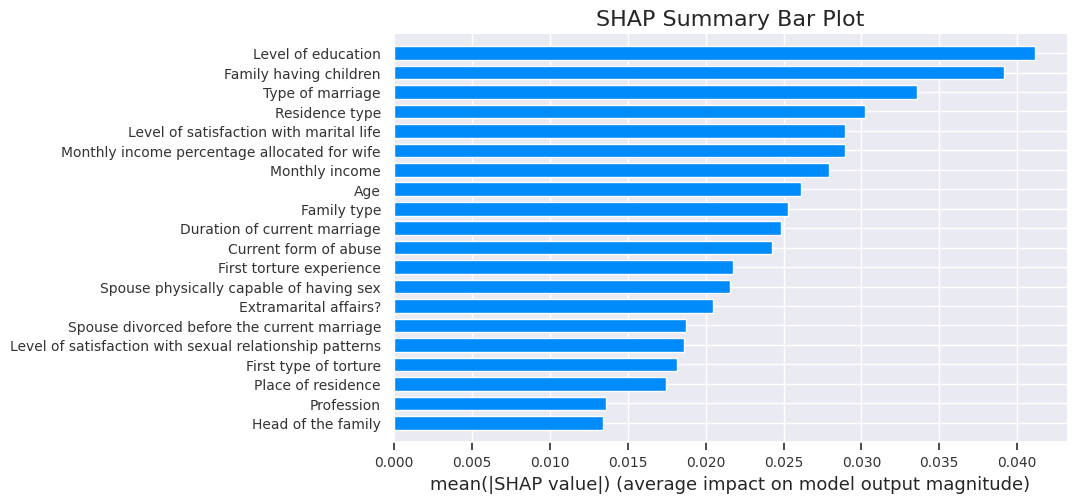}
        \caption{SHAP summary bar plot.}
    \end{subfigure}
    \begin{subfigure}{0.49\textwidth}
        \centering
        \includegraphics[width=\textwidth]{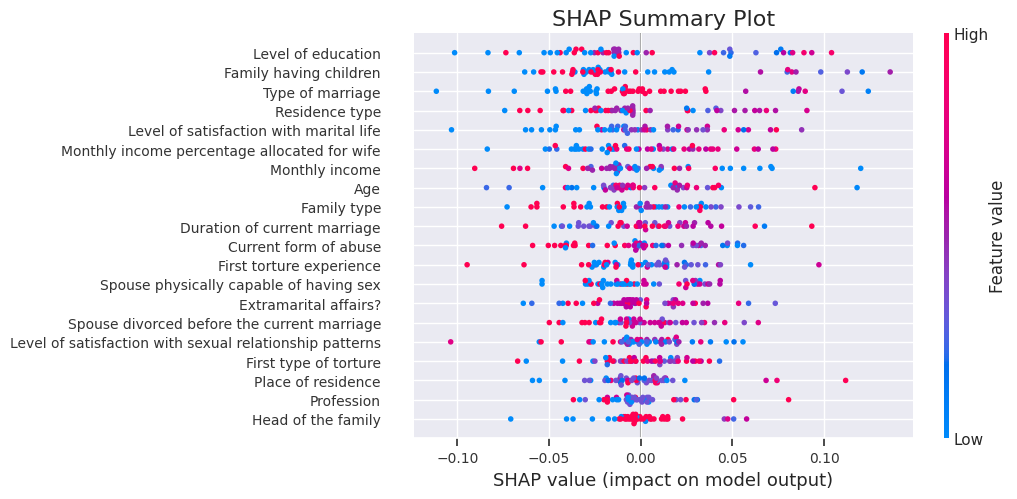}
        \caption{SHAP summary plot.}
    \end{subfigure}
    \caption{Dataset-level SHAP visualizations summarizing feature importance and their behavior.}
    \label{fig:dataset_summary}
\end{figure*}

Fig.~\ref{fig:dataset_summary} includes two complementary visualizations summarizing feature importance and behavior across the dataset. The summary bar plot highlights each feature's mean absolute SHAP values, indicating their overall impact on the model's predictions. Features like \textit{Level of education}, \textit{Family having children}, and \textit{Type of marriage} are identified as the most influential, while features such as \textit{Profession} and \textit{Head of the family} have relatively minor contributions. The summary plot provides a more granular view by combining feature importance with their directional impact. For example, high values of \textit{Monthly income} and \textit{Current form of abuse} tend to increase predictions, as indicated by the red dots, while low values of \textit{Type of marriage} decrease predictions, as shown by the blue dots. These visualizations offer a comprehensive dataset-wide understanding of feature relevance and behavior.

\begin{figure}[!ht]
\centering
\includegraphics[width=1\textwidth]{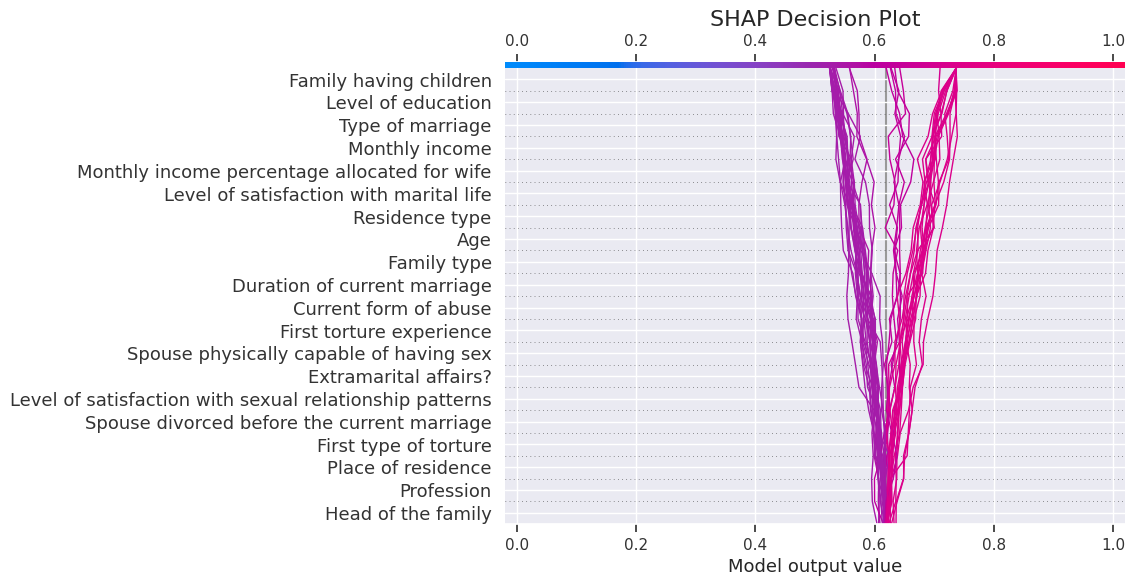}
\caption{SHAP decision plot showing aggregated feature impacts on predictions across instances.}
\label{fig:decision_plot}
\end{figure}

Fig.~\ref{fig:decision_plot} presents a SHAP decision plot, which visualizes the aggregated influence of features on the model's predictions across all instances. The x-axis represents the model's output values, while the y-axis ranks features by their importance. Each line corresponds to an instance, tracing the contribution of features to the final prediction. Features such as \textit{Family having children}, \textit{Level of education}, and \textit{Type of marriage} exhibit significant impacts, as evidenced by the steep lines in the plot. This visualization provides a high-level overview of how individual features interact to shape predictions consistently across the dataset.

\begin{figure}[!ht]
    \centering
    \begin{subfigure}{0.49\textwidth}
        \centering
        \includegraphics[width=\textwidth]{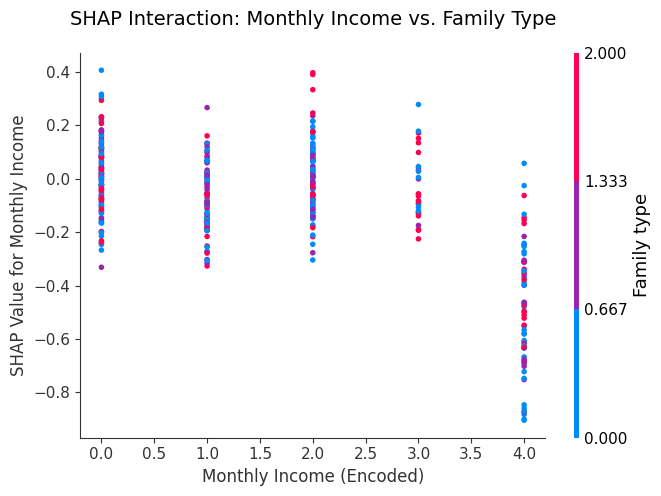}
        \caption{Interaction: Monthly Income $\times$ Family Type}
        \label{fig:shap_interactions_a}
    \end{subfigure}
    \begin{subfigure}{0.49\textwidth}
        \centering
        \includegraphics[width=\textwidth]{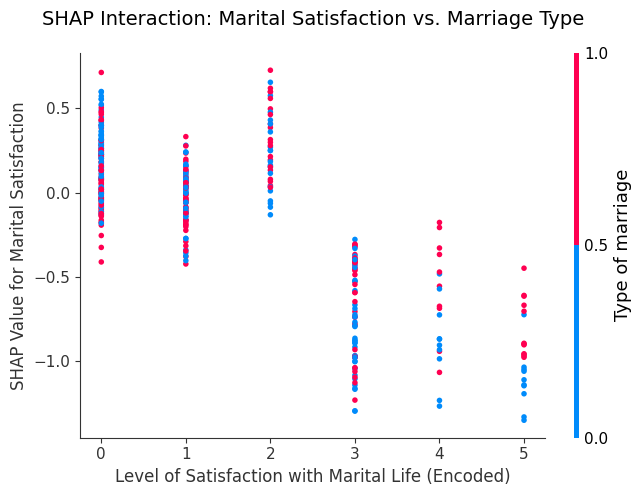}
        \caption{Interaction: Marital Satisfaction $\times$ Marriage Type}
        \label{fig:shap_interactions_b}
    \end{subfigure}
    \caption{SHAP dependence plots visualizing two key interaction effects learned by the stacking model.}
    \label{fig:shap_interactions}
\end{figure}

To provide deeper insights into the relationships within the data, we also conducted a SHAP interaction analysis to explore how combinations of features jointly influence MDV risk. The features for this analysis were chosen based on their high SHAP values, which indicated their significance in the model's predictions. Figure~\ref{fig:shap_interactions} displays two SHAP dependence plots for the most insightful interactions identified.  The first plot (Fig.~\ref{fig:shap_interactions_a}) shows that while low income is a general risk factor (positive SHAP value), this risk is significantly exacerbated in joint families. Conversely, the protective effect of high income is weaker in joint families compared to single-family households. The second plot (Fig.~\ref{fig:shap_interactions_b}) reveals that while marital dissatisfaction consistently increases risk, its negative impact is far more pronounced in arranged marriages than in love marriages. These findings confirm that our model did not just learn simple, linear effects but successfully captured the complex, combinatorial nature of MDV risk.

\subsubsection{LIME Interpretations}
\begin{figure}[!ht]
    \centering
    \includegraphics[width=1\textwidth]{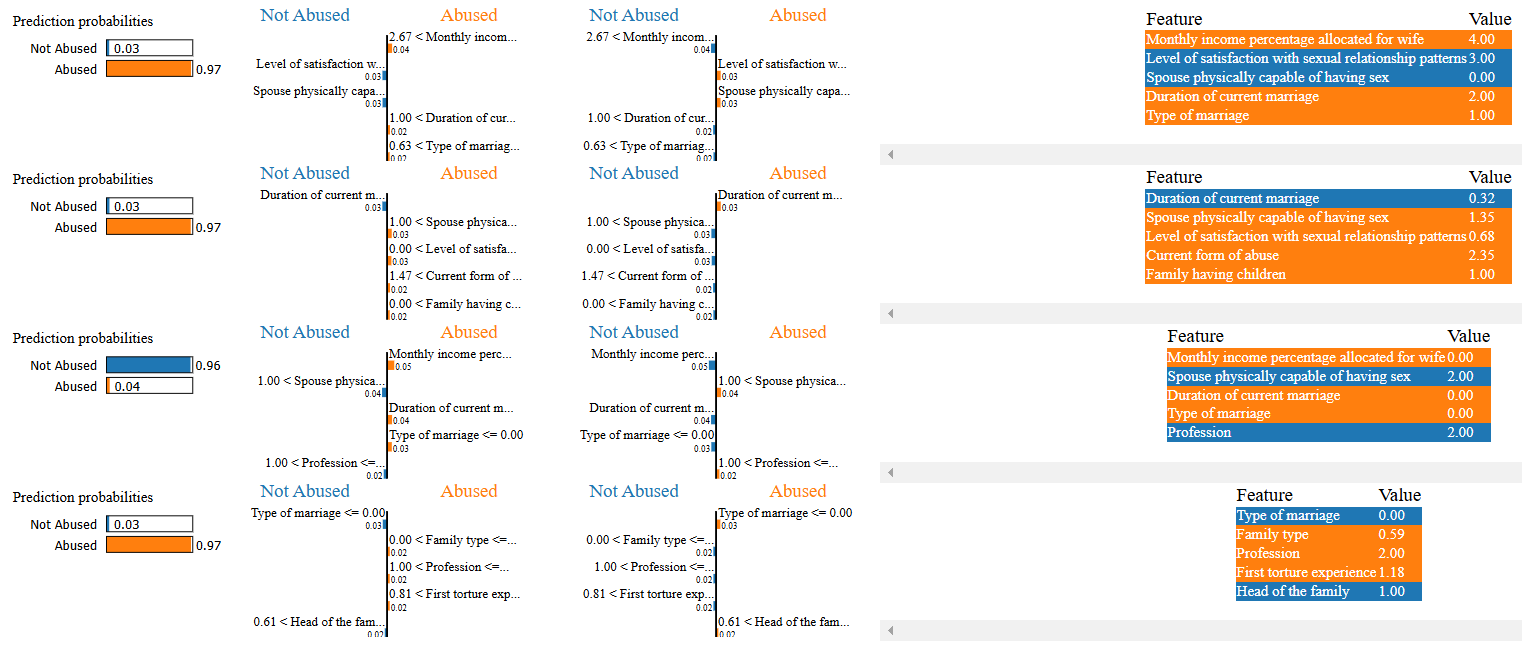}
    \caption{LIME visualization explaining predictions for multiple instances.}
    \label{fig:lime_analysis}
\end{figure}

LIME provides localized insights into the predictions made by the proposed model (ANN-CatBoost stacking + LR). To make these abstract predictions tangible and demonstrate their practical utility, we can interpret the LIME outputs as illustrative case studies. These are not based on qualitative interviews but are data-driven narratives constructed from the model's explanations, showing how different risk profiles lead to a prediction. Fig.~\ref{fig:lime_analysis} illustrates the feature contributions for four such instances.

\paragraph{Illustrative Case 1 (Financial Strain)} The first instance is predicted as ``Abused'' with 97\% probability. The feature contributing most strongly to this prediction is the \textit{monthly income percentage allocated for wife}, suggesting a narrative where financial arrangements are a significant source of conflict or control. Conversely, the model identifies the \textit{level of satisfaction with sexual relationship patterns} as a minor protective factor pushing against the prediction. An actionable insight from this specific profile is the need for interventions focused on financial counseling or economic stress management, rather than a one-size-fits-all approach. 

\paragraph{Illustrative Case 2 (Active Abuse)} The second instance is predicted as ``Abused'' with 97\% probability, but it points to a different narrative. Here, the prediction is driven by features indicating an active and ongoing problem: \textit{Current form of abuse} and the partner's physical capability. The model's focus on these features provides a clear insight for intervention: this individual likely requires immediate support focused on safety planning and crisis management. Interestingly, the \textit{duration of current marriage} slightly reduces the risk score, suggesting that newer marriages might be a critical window for early-stage preventive programs. 

\paragraph{Illustrative Case 3 (Protective Factors)} In contrast, the third instance represents a low-risk profile, with a 96\% probability for ``Not Abused." The key protective factors identified by LIME are \textit{spouse physically capable of having sex} and \textit{profession}. This narrative suggests a stable relationship where factors like mutual health and professional stability may act as buffers against abuse. The neutral contribution of other features supports the idea of a predictable dynamic, reinforcing that interventions for such profiles might focus on maintaining relational health. 

\paragraph{Illustrative Case 4 (Complex Risk Pathway)} The final instance, also predicted as ``Abused,'' highlights a more complex risk pathway. Here, the prediction is driven by \textit{profession} and \textit{first torture experience}. This suggests a narrative where profession-related stressors (e.g., unemployment, working abroad) combined with an early history of abuse are the primary drivers of risk.

\subsection{Practical Applications and Responsible Use}
While our model demonstrates high predictive accuracy, its transition from a research finding to a real-world tool requires careful consideration of both its practical applications and its inherent ethical risks. The primary value of this model is not as an automated decision-maker but as a supportive tool to augment the expertise of professionals in various fields. Key applications include:
\begin{enumerate}
    \item \textbf{Social Work and Counseling:} Caseworkers could use the model as a risk-assessment aid during intake to help prioritize cases and identify non-obvious risk factors. The XAI outputs, such as the LIME interpretations in our study, can guide conversations by highlighting specific areas of concern (e.g., financial strain, a history of early abuse), allowing for more personalized support plans.

    \item \textbf{Public Policymaking:} The model's global feature importance analysis provides quantitative evidence for policy design. If financial dependency is consistently identified as a top predictor of MDV in a region, it offers a strong justification for funding economic empowerment programs and support services as a targeted domestic violence prevention strategy.

    \item \textbf{Law Enforcement Training:} The model should \textbf{not} be used for predictive policing or as evidence against individuals. However, its insights can be integrated into training curricula to help officers better recognize the complex and often subtle combinations of factors that contribute to domestic violence risk, moving beyond simplistic stereotypes.
\end{enumerate}

The deployment of a predictive model in such a sensitive domain also carries significant ethical responsibilities that must be proactively managed. The primary risks include:
\begin{enumerate}
    \item \textbf{Misclassification:} A false positive (incorrectly flagging a low-risk individual) could lead to unnecessary stigma, while a false negative (failing to identify a true victim) could have tragic, life-threatening consequences. To mitigate this, the model's output must always be treated as a supplemental insight within a “human-in-the-loop” system where a trained professional makes the final judgment.

    \item \textbf{Misuse and Bias:} Using the model for punitive actions like creating \textquotedblleft watch lists\textquotedblright would be a gross misuse of its purpose. Furthermore, applying the model to new populations without retraining could amplify existing societal biases. Therefore, any potential deployment must be bound by strict ethical guidelines limiting its use to prevention and support, and must include a protocol for retraining and bias auditing before being applied to new demographics.
\end{enumerate}

\section{Conclusions}
\label{sec:Conclusion}
This study presents a novel computational framework to investigate the complex and often overlooked issue of MDV in Bangladesh. By implementing stacking ensemble ML and EDA, we addressed the inherent challenges of categorical data imbalance and scarcity to identify critical patterns of abuse. Our investigation, based on survey data from nine major Bangladeshi cities, confirms the multifaceted nature of MDV, where factors such as financial dependency, marital dissatisfaction, and socio-cultural dynamics like joint family structures and arranged marriages are significant determinants of victimization. The comparative evaluation of 15 models demonstrated that ensemble techniques consistently outperform traditional methods. The proposed ANN-CatBoost stacking model, in particular, yielded the highest performance, with a 95\% accuracy and a 99.29\% AUC. This model proved robust in navigating a highly categorical and imbalanced dataset, effectively balancing sensitivity and specificity. XAI techniques further validated the utility of this approach. Both model-specific feature importance and model-agnostic methods like SHAP and LIME showed the significance of demographic, socio-economic, and relational variables, such as level of education, family structure, and type of marriage, in predicting MDV. A key contribution of this research is demonstrating the superiority of ensemble ML over traditional statistical methods for this specific problem. This was shown by the contrast between our initial exploratory analysis and the final model's performance. Our $\chi^2$ analysis failed to find significant bivariate associations between victimhood and key indicators like education or profession. On the other hand, the feature importance and SHAP analyses of our proposed ANN-CatBoost model identified these very features as critical predictors. This disparity confirms that MDV in this context is not affected by a single factor, but by the sophisticated interplay of numerous socio-economic, demographic, and relational variables. Our work shows that to predict and understand MDV effectively, we must move beyond simple associations and embrace models that can capture the true complexity of the phenomenon. While the current model's feature weights reflect the unique socio-cultural context of Bangladesh, the methodological framework itself, combining stacking ensemble learning with XAI for interpretability, is highly adaptable. To adapt this work for a different country or cultural context, researchers would need to collect local data and retrain the model. This process would not only serve as a universal template for future research but would also allow for powerful cross-cultural comparisons of MDV risk factors.

Although this study provides valuable insights, its limitations must be acknowledged. The reliance on survey data from major urban centers means the findings may not fully capture the diversity of experiences in rural areas, and the focus on professionals could introduce sampling bias. Also, reliance on self-reported data, even when collected anonymously, presents inherent challenges related to underreporting or social desirability bias, particularly given the deeply entrenched stigma surrounding male victimhood. Finally, it is essential to note that while our LIME analysis provides illustrative case studies, this quantitative approach is not a substitute for actual qualitative research; a dedicated study using in-depth interviews is a critical next step to capture the lived experiences behind the data. So, future research should aim to expand the scope to include rural and diverse socio-economic populations. Qualitative methodologies, such as in-depth interviews, are recommended to provide a deeper understanding of the lived experiences of male victims. Finally, longitudinal studies are needed to track the dynamics of MDV over time, while the application of natural language processing (NLP) to analyze online narratives presents a promising avenue for developing real-time, preventative strategies. Such continued investigation is essential to address the remaining gaps in our MDV understanding and identify new risk factors.

\section*{Ethical Approval and Accordance } 
This study was approved by the Office of the Director, Research \& Extension Center, Khulna University of Engineering \& Technology, Khulna, Bangladesh \texttt{KUET/DRE/2023/27(1)}. All procedures involving human participants were conducted in accordance with institutional and national ethical standards and complied with the Declaration of Helsinki and its later amendments or comparable ethical standards.

\section*{Consent to Participate} 
Informed consent was obtained from every participant after full disclosure of the study’s aims, procedures, potential risks, and benefits. Participation was voluntary, and participants had the right to withdraw at any time without consequence.

\section*{Declarations}
\subsection*{Data Availability}
% The data used in this study are available upon reasonable request. All datasets and related resources will be made publicly available upon acceptance of the manuscript.
The dataset supporting the findings of this study is available at the Mendeley Data repository: \href{https://doi.org/10.17632/97xnx8nf22}{https://doi.org/10.17632/97xnx8nf22}

\subsection*{Funding}
The authors declare that no funds, grants, or other support were received during the preparation of this manuscript.

\subsection*{Competing Interests}
The authors have no relevant financial or non-financial interests to disclose.

\subsection*{Author Contributions Statement}
\textbf{M.A.J.}: Conceptualization, Methodology, Data curation, Writing -- Original Draft Preparation, Software, Visualization, Investigation.\\
\textbf{S.A.N.}: Writing -- Original Draft Preparation, Formal Analysis, Software.\\
\textbf{F.T.J.L.}: Writing -- Original Draft Preparation.\\
\textbf{M.F.M.}: Supervision, Project Administration, Writing -- Review \& Editing.\\
\textbf{M.J.H.}: Writing -- Review \& Editing.

\subsection*{Clinical Trial Number} 
Not applicable.

\subsection*{Consent to Publish declaration} 
Not applicable. No individual identifying data is included in this publication.

\bibliographystyle{apacite}
\bibliography{main}

\end{document}